\documentclass[12pt,preprint]{aastex}
\usepackage{natbib}
\usepackage{graphicx}
\begin{document}
\title{Green Pea Galaxies and cohorts: Luminous Compact Emission-Line Galaxies in the Sloan Digital Sky Survey}
\author{Yuri I. Izotov}
\affil{Main Astronomical Observatory, Ukrainian National Academy of Sciences,
27 Zabolotnoho str., Kyiv 03680, Ukraine}
\email{izotov@mao.kiev.ua}
\author{Natalia G. Guseva}
\affil{Main Astronomical Observatory, Ukrainian National Academy of Sciences,
27 Zabolotnoho str., Kyiv 03680, Ukraine}
\email{guseva@mao.kiev.ua}
\and
\author{Trinh X. Thuan}
\affil{Astronomy Department, University of Virginia, P.O. Box 400325, 
Charlottesville, VA 22904-4325}
\email{txt@virginia.edu}

\begin{abstract}
We present a large sample of 803 star-forming luminous compact galaxies (LCGs) 
in the redshift range $z$ = 0.02 -- 0.63, selected from Data Release 7
of the Sloan Digital Sky Survey (SDSS). The global properties
of these galaxies are similar to those of the so-called ``green pea'' 
star-forming galaxies, in the redshift range $z$ = 0.112 -- 0.360 and  
selected from the SDSS on the basis of their 
green color and compact structure. In contrast to green pea galaxies, 
our LCGs are selected on the basis of 
both their spectroscopic and photometric properties, 
resulting in a $\sim$ 10 times larger sample, with galaxies spanning a 
redshift range $\ga$ 2 times larger.
We find that the oxygen abundances 
and the heavy element abundance ratios in LCGs do not differ from
those of nearby low-metallicity blue compact dwarf (BCD) galaxies.
The median stellar mass of LCGs is $\sim$ 10$^9$ $M_\odot$. However,
for galaxies with high EW(H$\beta$), $\ge$ 100\AA, it is only 
$\sim$ 7$\times$10$^8$ $M_\odot$.
The star formation rate in LCGs varies in the large 
range of 0.7 -- 60 $M_\odot$ yr$^{-1}$, with a median value of 
$\sim$ 4 $M_\odot$ yr$^{-1}$, a factor of $\sim$ 3 lower than
in high-redshift star-forming galaxies at $z$ $\ga$ 3.
The specific star formation rates in LCGs are extremely high and vary in
the range $\sim$ 10$^{-9}$ -- 10$^{-7}$ yr$^{-1}$, comparable 
to those derived in high-redshift galaxies. 
\end{abstract}

\keywords{galaxies: abundances --- galaxies: irregular --- 
galaxies: ISM --- galaxies: star formation --- H {\sc ii} regions}

\section{Introduction}

Recently, \citet{C09} reported on a new class of luminous compact galaxies at 
redshifts $z$ = 0.112 -- 0.360, discovered by the Galaxy Zoo project from the 
Sloan Digital Sky Survey (SDSS) images. The Galaxy Zoo project collected 
simple morphological classifications of nearly 900 000 galaxies drawn from the 
SDSS, contributed by hundreds of thousands of volunteers \citep{L08,L11}. 
In particular, the attention was drawn on compact objects named 
``green pea'' galaxies.
Their appearance is 
mainly caused by the very strong [O {\sc iii}] $\lambda$5007\AA\ 
optical emission line, with an unusually large equivalent width
of up to $\sim$ 1000 \AA, redshifting into the SDSS $r$ band, resulting in 
a green color in the $gri$ composite SDSS images 
(green being the color represented by the $r$ band in these composite images). 
 \citet{C09} discussed a well-defined sample 
of 251 color-selected objects located mainly in low-density regions. They
found that the median environmental density around the green pea galaxies
is less than two-thirds of that around normal galaxies.  
Most of the green pea galaxies are strongly star-forming, although 
there are some active galactic nuclei interlopers, including eight newly 
discovered narrow-line Seyfert 1 galaxies. 
For a further spectroscopic analysis, \citet{C09} selected a subsample of 80
star-forming galaxies and found that they are
low-mass galaxies ($M$ $\sim$ 10$^{8.5}$ -- 10$^{10}$ $M_\odot$), with high 
star formation rates ($\sim$ 10 $M_\odot$ yr$^{-1}$) and solar metallicities, 
if the solar calibration of  
 \citet{As09}, 12 + log (O/H)$_\odot$ = 8.7, 
is adopted. \citet{C09} found that green 
pea galaxies have some of the highest specific star formation rates seen in 
the local Universe (up to 10$^{-8}$ M$_\odot$ yr$^{-1}$), and are similar in size, mass, luminosity and metallicity 
to luminous blue compact galaxies. They are also similar to high-redshift 
ultraviolet-luminous galaxies (Lyman-break galaxies and Ly-$\alpha$ emitters).
Those authors suggested that the green pea galaxies
may be the last remnants of a mode of star formation common in
the early Universe, 
and therefore provide an excellent local laboratory for  
studying the extreme star formation 
processes that occur in high-redshift galaxies. 

Later, \citet{A10} investigated the oxygen and nitrogen chemical 
abundances in the green pea galaxies of \citet{C09}. 
They found these systems to be  
genuine metal-poor galaxies, with mean oxygen abundances 
12 +log O/H $\sim$8.0, or 0.2 solar. 
While the N/O ratios of green pea galaxies 
follow the relation with stellar mass of local 
star-forming galaxies, \citet{A10} find that the green pea mass-metallicity relation is offset by $\sim$ 0.3 dex to lower metallicities. 
They argue that recent interaction-induced inflow of gas, possibly coupled 
with a selective metal-rich gas loss, driven by supernova winds, may explain 
their findings. 
Interaction-induced infall of gas would also 
be consistent with the known galaxy properties such as high specific star 
formation rates, extreme compactness, and disturbed optical morphologies. 
\citet{A10} concluded that the green pea galaxy properties seem to be not 
common in the nearby universe, suggesting that 
the green pea phase is a short and extreme stage of their
evolution. They agreed with \citet{C09} that 
these galaxies may allow to study in great detail 
many processes, 
such as starburst activity and chemical enrichment, under 
physical conditions approaching those in galaxies at higher redshifts.

In this paper, we wish to further the studies of green pea galaxies 
and reexamine the above conclusions of \citet{C09} and \citet{A10}
 by  
studying a considerably larger sample of luminous compact emission-line
galaxies selected 
from the Data Release 7 (DR7) of the Sloan Digital Sky Survey 
\citep{A09}.
We adopt an entirely different approach in the selection of 
our galaxy sample. 
 In contrast to the study of \citet{C09}, our luminous 
compact galaxies are selected using spectroscopic criteria which are
supplemented by morphological criteria.
The selection criteria
of our galaxies are described in Section 2. Their element abundances are 
derived in Section 3. The luminosity-metallicity relation
for the sample galaxies is discussed in Section 4.
In Section 5 we describe a technique for 
galaxy stellar mass determination, taking into account the gaseous continuum 
emission. We derive stellar masses for the young and old 
stellar populations in each galaxy. In Section 6 we compare the physical
properties of 
starbursts in luminous compact galaxies.
The mass-metallicity relation for the sample galaxies is
derived in Section 7. In Sections 8 and 9, we examine respectively 
their star formation rates 
and the age
of their stellar populations. Our conclusions are presented in Section 10.

\section{A sample of luminous compact galaxies}

\subsection{Selection criteria}

The photometric selection criteria used by \citet{C09}, based 
on the green appearance of a galaxy in the SDSS images, 
pick out galaxies in a relatively
narrow range of redshifts, $z$ $\sim$ 0.1 -- 0.3: those redshifts are 
such as to make the strong H$\beta$, [O {\sc iii}] $\lambda$4959\AA\ and
[O {\sc iii}] $\lambda$5007\AA\ optical emission lines fall in the SDSS $r$ 
band. Such an approach fails to find galaxies with similar properties at lower 
and higher redshifts because their colors would then not be green. 

To avoid this restriction, we adopt here a different approach. 
We select luminous compact emission-line 
galaxies (LCGs) by using criteria that are based on both  
spectra and images in the SDSS DR7. These criteria are as follows:

1) the extinction-corrected 
luminosity of the H$\beta$ emission line is 
greater than $L$(H$\beta$) = 3$\times$10$^{40}$ erg s$^{-1}$. 
The extinction
coefficient is derived from the observed Balmer decrement in the SDSS spectra.

\begin{figure*}
\figurenum{1}
\hbox{\includegraphics[angle=-90,width=1.0\linewidth]{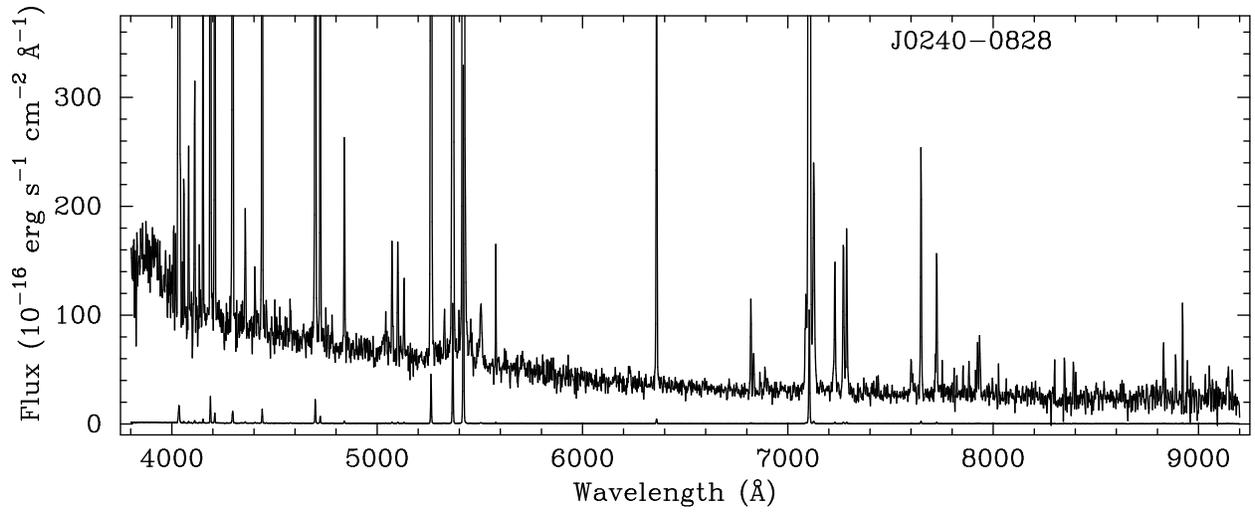} 
}
\figcaption{SDSS spectrum of a luminous compact emission-line galaxy in  
our sample obtained with the 2.5m Apache Point Observatory Sloan 
telescope in a 1h exposure. 
The ordinate scale is for the lower spectrum. The upper 
spectrum is the lower spectrum expanded by a factor of 100 along the ordinate. 
Weak emission lines with fluxes of $\sim$ 
1\% that of H$\beta$ are clearly detected 
in this spectrum. \label{fig1}}
\end{figure*}

2) the equivalent
width of the H$\beta$ emission line is high, EW(H$\beta$) $\ga$ 50 \AA. 
This criterion helps to 
select only objects with strong emission lines in their 
spectra (Fig. \ref{fig1}), and thus containing young starbursts with 
ages 3 - 5 Myr.
We note that most of the green pea galaxies 
from \citet{C09} have EW(H$\beta$) $\ga$ 100 \AA. Thus, our sample of LCGs
covers a larger range of EW(H$\beta$). 

Contrary to \citet{C09}, we do 
not use the equivalent width of the [O {\sc iii}] $\lambda$5007 emission line 
as a selection criterion. This is because the strength of the [O {\sc iii}]
$\lambda$5007 emission line in young starbursts is metallicity-dependent.
It is strongest at 12+logO/H $\sim$ 8.0, decreasing at both higher and 
lower O abundances. Therefore, using the strength    
the [O {\sc iii}] $\lambda$5007 emission line as a selection 
criterion would unduly bias the sample toward galaxies with a narrow range of  
O abundances near 12+logO/H $\sim$ 8.0. On the other hand, 
the use of the H$\beta$ emission line allows us to pick out 
galaxies with a considerably larger range of O abundances. 

3) only galaxies with a well-detected 
[O {\sc iii}] $\lambda$4363 \AA\ emission line in their spectra, with a 
flux error less than 50\% of the line flux, are selected. This criterion
allows an accurate abundance determination using the direct method, 
in contrast to \citet{C09} who used the less accurate empirical strong-line 
method for abundance determination. If we do not 
impose criterion 3, but only criteria 1 and 2,  
then the number of selected galaxies would increase
only by $\sim$20\%. 
Thus, when criterion 3 is imposed, the decrease in the 
number of galaxies is small enough so as 
not to introduce large biases in the physical properties of the 
selected sample. On the other hand, the benefit of being able to  
determine accurate O abundances and study trends with metallicity is 
inestimable.

4) galaxies with obvious evidence of Seyfert 2 spectral features, such as 
strong [Ne {\sc v}] $\lambda$3426, He {\sc ii} $\lambda$4686, 
[O {\sc i}] $\lambda$6300, and [S {\sc ii}] $\lambda$6717, $\lambda$6731 
emission lines are
excluded. Thus, we select only star-forming galaxies.

5) on their SDSS images, galaxies are nearly compact at low redshifts, 
and unresolved at high redshifts.

Using these criteria, we constructed 4 subsamples of LCGs: 
1) subsample 1 includes 161 galaxies with EW(H$\beta$) $\ge$ 100\AA\ and
a round shape, without evident signs of disturbed morphology. 
Thus, this subsample 
should be most similar in properties to the star-forming green pea galaxy 
sample of \citet{C09};  
2) subsample 2 consists of 159 
galaxies with EW(H$\beta$) $\ge$ 100\AA\ and with some sign of disturbed 
morphology suggesting interaction and the presence of 
more than one star-forming region; 
3) subsample 3 is the same as subsample 1, but it consists of 116 galaxies 
with a lower EW(H$\beta$), covering the range 
100\AA\ $>$ EW(H$\beta$) $\ge$ 50\AA; and 4) subsample 4 is the same as 
subsample 2, but it consists of 367 galaxies with a lower EW(H$\beta$),
100\AA\ $>$ EW(H$\beta$) $\ge$ 50\AA. The galaxies from subsample 4
often show two or more compact knots of star formation.
Thus, the total number of selected galaxies is 803, or $\sim$ 10 times larger 
than the sample of star-forming green pea galaxies of \citet{C09}. 
In general, the selected LCGs are faint objects with low extinction.
Their median SDSS $g$
apparent magnitude and reddening $E$($B-V$) range from 20.3 and 0.10 
for subsample 1 to 18.4 and 0.15 for subsample 4 (Table \ref{tab1}).

\subsection{A Luminosity Upper Limit}

\begin{figure*}
\figurenum{2}
\hbox{\includegraphics[angle=-90,width=1.0\linewidth]{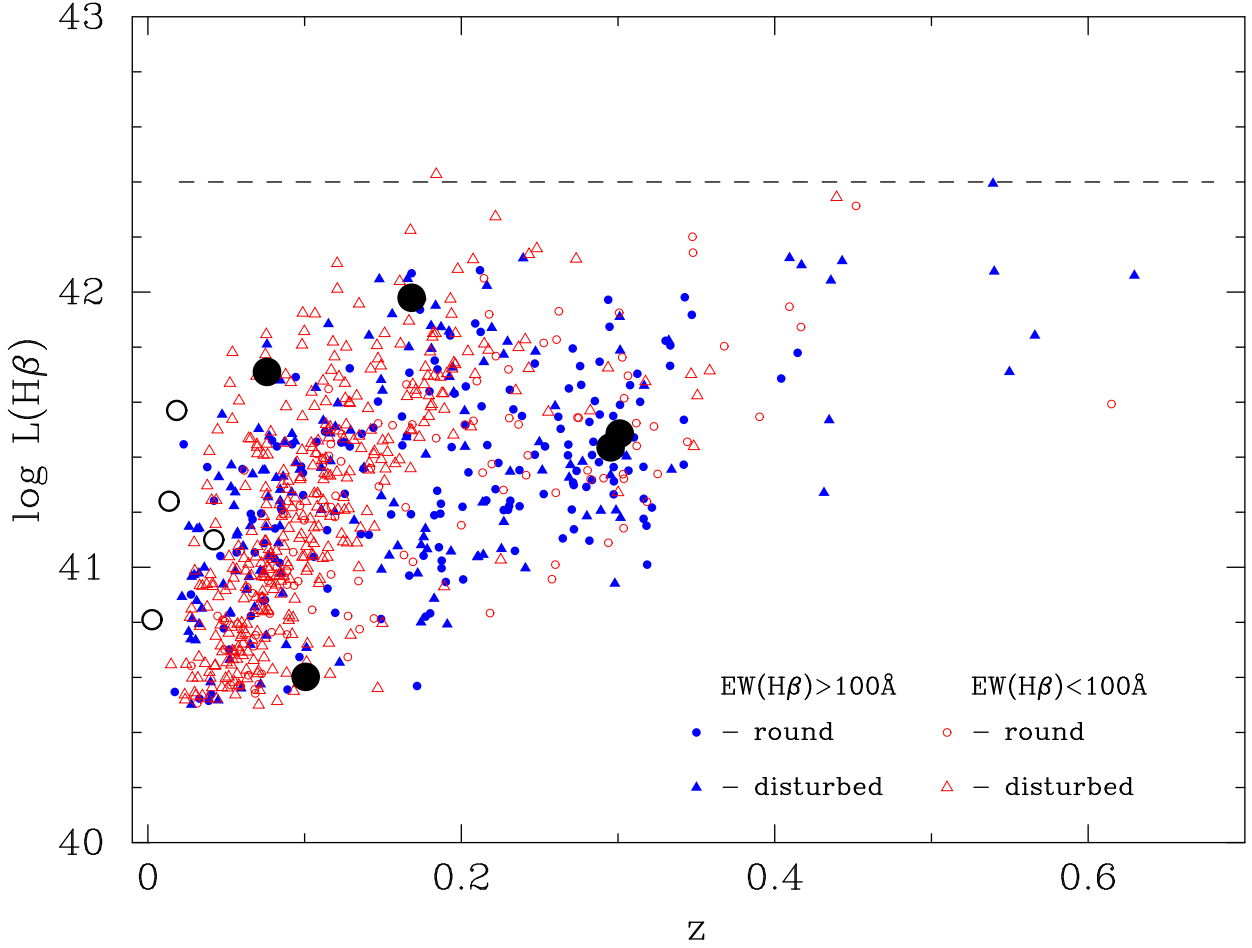} 
}
\figcaption{Extinction-corrected
H$\beta$ luminosity $L$(H$\beta$) vs. redshift $z$ for the LCG sample. LCGs 
with EW(H$\beta$) $\ge$ 100\AA\ are shown by filled symbols.  
Those with a round shape and those with some signs of disturbed morphology
are shown respectively by small filled circles and small filled triangles.
LCGs with 50\AA\ $\le$ EW(H$\beta$) $<$ 100\AA\ are shown by open 
symbols. Those with a round shape and those with some signs of 
disturbed morphology
are shown respectively by small open circles and small open triangles.
Well-studied local  
LCGs, also known as Blue Compact Dwarf galaxies, such as 
Mrk 930, SBS 0335--052E, HS 0837+4717 and II Zw 40
are shown by large open circles. Low-metallicity AGN
candidates from \citet{IT08} and \citet{I10} are shown by large filled
circles. \label{fig2}}
\end{figure*}

The distribution of the H$\beta$ luminosity $L$(H$\beta$) with redshift $z$
is shown in Fig. \ref{fig2}. It is seen that the selected galaxies
span a large range of redshifts, from $\sim$ 0.02 to $\sim$ 0.63.
The apparent increase of $L$(H$\beta$) with redshift, for $z$$\la$0.2,  
is likely the result of a   
selection effect (called the Malmquist bias): only 
the brightest galaxies at high
redshift were observed spectroscopically in the SDSS.  
Fig. \ref{fig2} shows that galaxies with 
EW(H$\beta$) $\ge$ 100\AA\ are distributed
more uniformly with redshift than those with lower EW(H$\beta$), which 
tend to populate the lower redshift region. 
This difference is likely also due to 
a selection effect. The [O {\sc iii}] $\lambda$4363\AA\ emission line
in objects with a high EW(H$\beta$) is stronger. Hence it can be measured with
a good accuracy even in high redshift objects, 
which helps to fulfill the selection criterion 3 above 
and allows the inclusion of the galaxy in the LCG sample.

Fig. \ref{fig2} shows a remarkable feature: there is
an upper limit of $L$(H$\beta$) at the level of $\sim$ 2.5$\times$10$^{42}$
erg s$^{-1}$ (shown by the horizontal dashed line),
which is almost not redshift dependent,  
despite the fact that none of our selection criteria sets an upper limit
to the H$\beta$ luminosity. The presence of this upper limit  
suggests the existence of a self-regulating mechanism in star formation 
which somehow forbids the formation of more luminous star-forming regions.
There is probably a negative feedback mechanism due to 
the intense UV radiation
of the present starburst which prevents further star formation in LCGs. While
beyond the scope of the present paper, the role of the ionizing radiation
from young massive starbursts in shutting off star formation 
beyond a given threshold luminosity, should be investigated.

\subsection{Blue compact dwarf, green pea and luminous compact galaxies: 
a continuum in luminosity}

The range of $z$ in Fig. \ref{fig2} is $\sim$ 2.5 times larger than that for 
the green pea galaxies of \citet{C09}. 
Those authors and \citet{A10} suggested that green pea galaxies form 
a new class of objects. We will argue now that this is not the case and that 
they are just a subset of LCGs. 
Nearby ($z$ $\sim$0) luminous compact emission-line
galaxies have been known for decades. 
Examples of such objects are, 
in order of decreasing $L$(H$\beta$) (we give the 
logarithm of the aperture- and 
extinction-corrected $L$(H$\beta$) in parentheses for each object in 
units of erg s$^{-1}$): 
Mrk 930 (41.57), SBS 0335--052E (41.24), HS 0837+4717 (41.10) 
and II Zw 40 (40.81). Some of them have been studied in great detail
\citep[see e.g. the studies of SBS 0335--052E by ][]{I90,I97,T97,TI05}. 
These objects are known as 
blue compact dwarf galaxies (BCD) and are plotted as large open circles in 
Fig. \ref{fig2}. A recent review of BCD properties is given by \citet{T08}.
We see that the $L$(H$\beta$) of these BCDs are comparable to those of  
LCGs at the low-luminosity end of the sample.
Other well-known BCDs have lower $L$(H$\beta$), below our cut-off: 
Tol1214--277 (40.43), I Zw 18 (39.72 for a distance of 15 Mpc) and Mrk 209 
(39.55). In other words, LCGs are the same type of objects as BCDs except that 
they are more luminous in the mean. 

\begin{figure*}
\figurenum{3}
\hbox{\includegraphics[angle=0,width=0.25\linewidth]{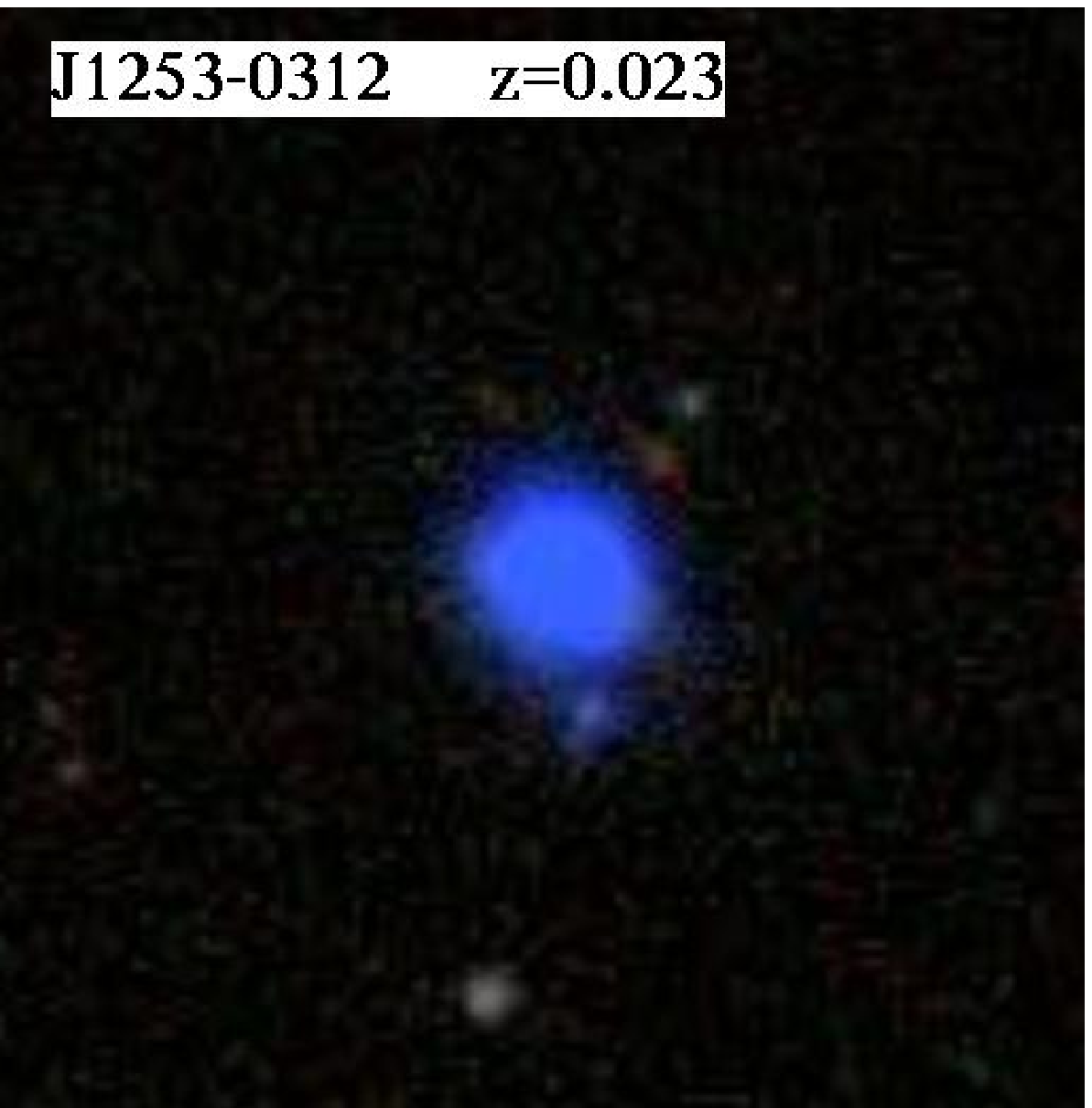} 
\includegraphics[angle=0,width=0.25\linewidth]{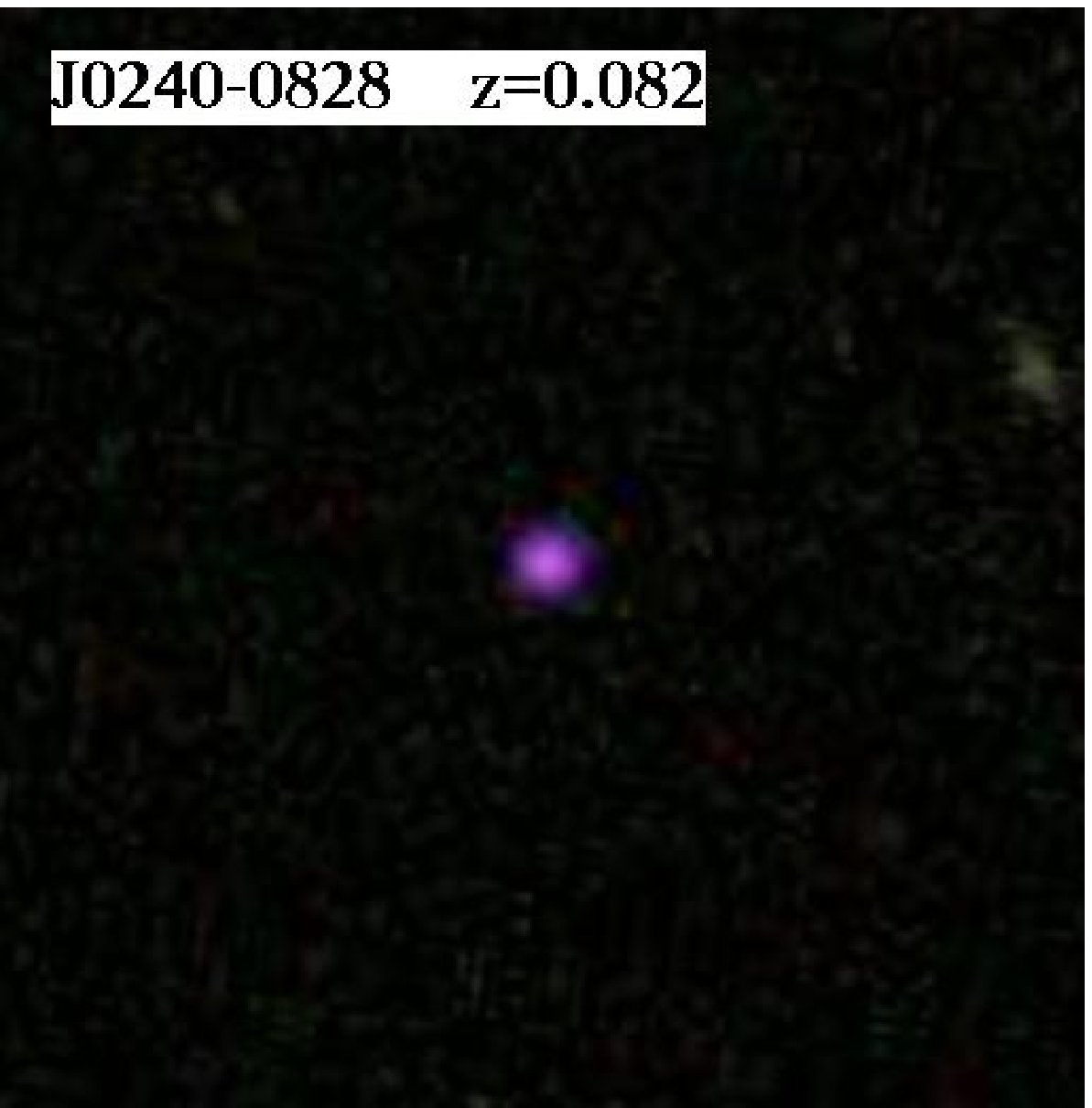} 
\includegraphics[angle=0,width=0.25\linewidth]{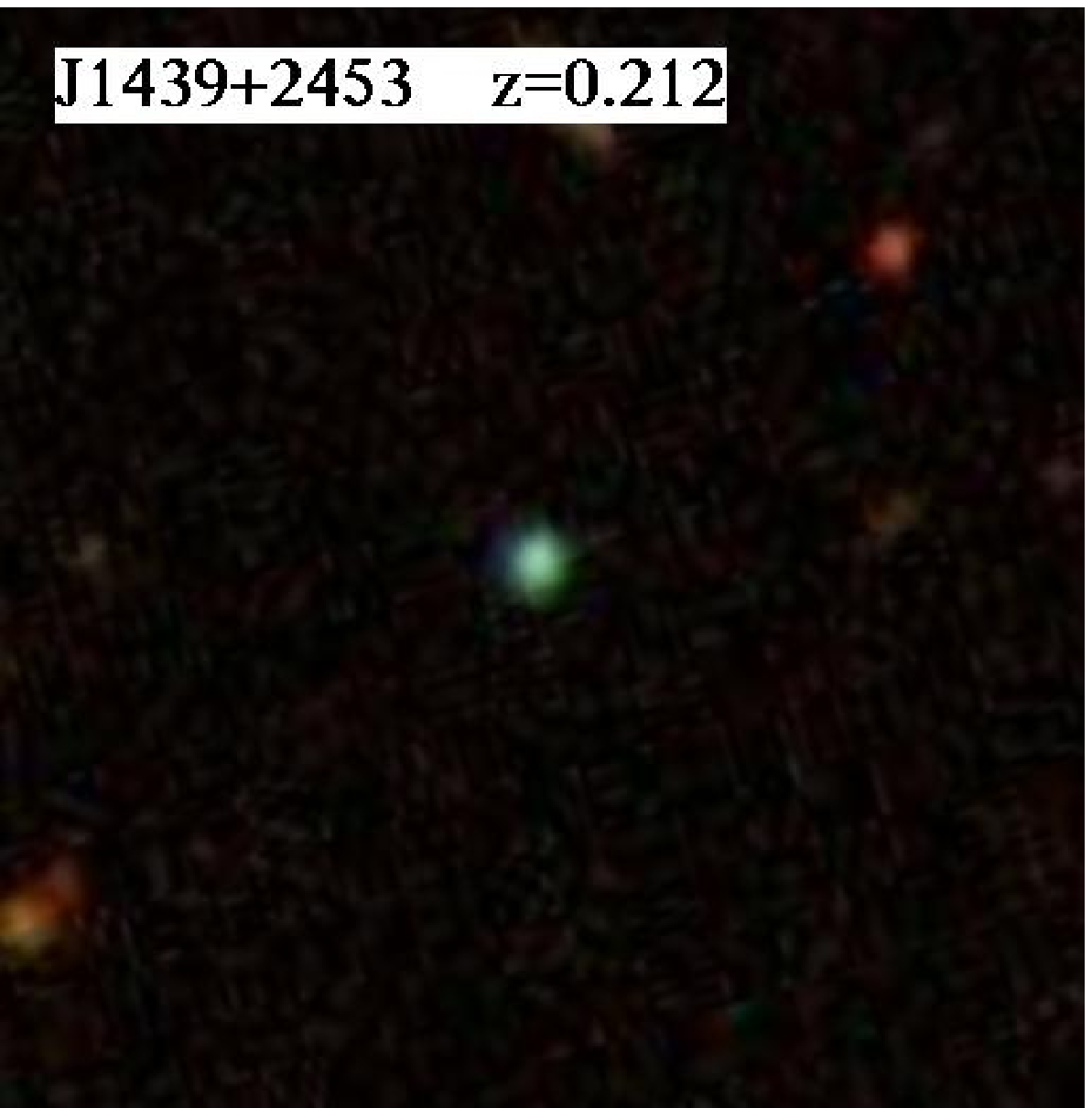} 
\includegraphics[angle=0,width=0.25\linewidth]{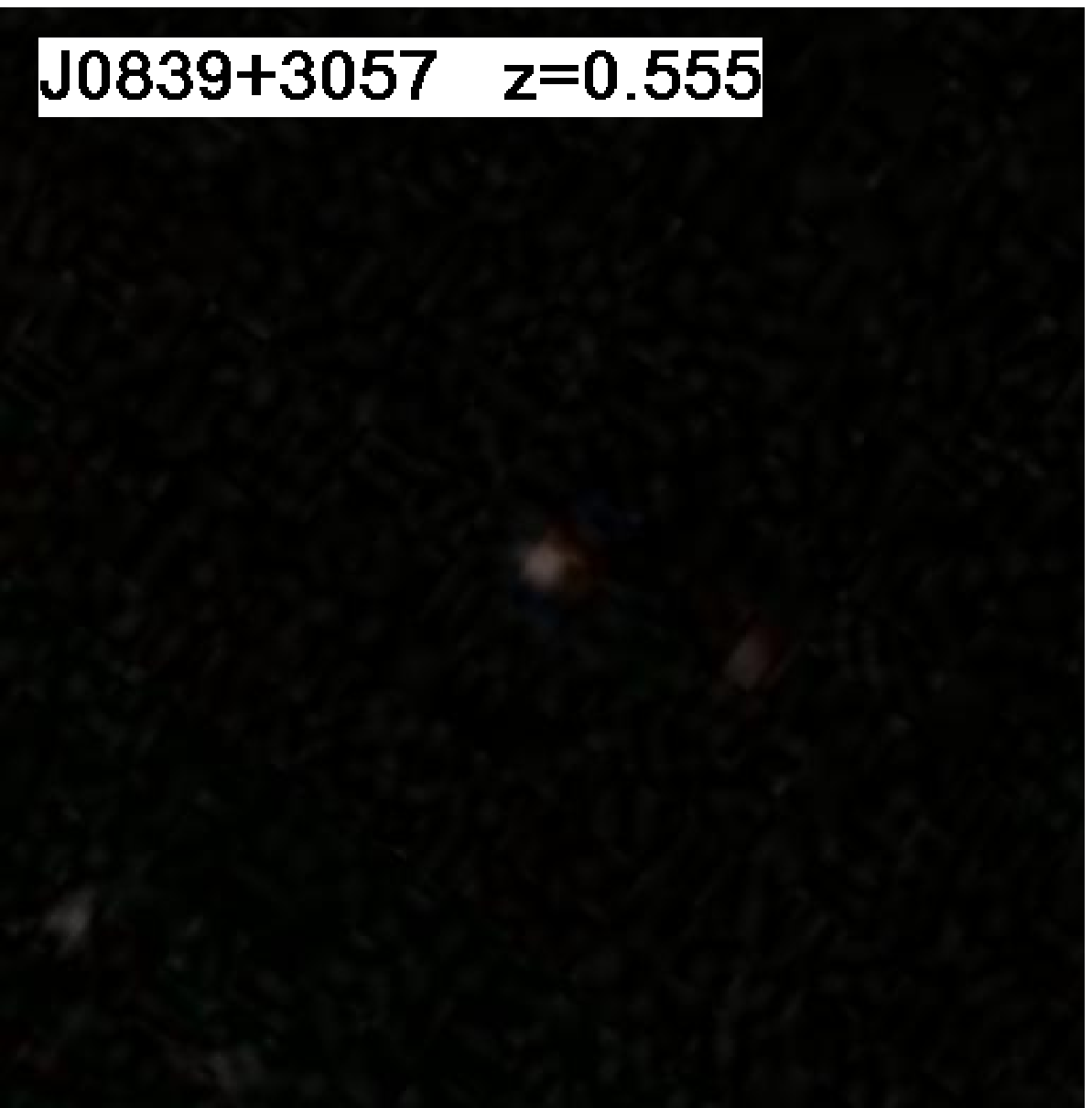}} 
\vspace{0.1cm}
\hbox{\includegraphics[angle=0,width=0.25\linewidth]{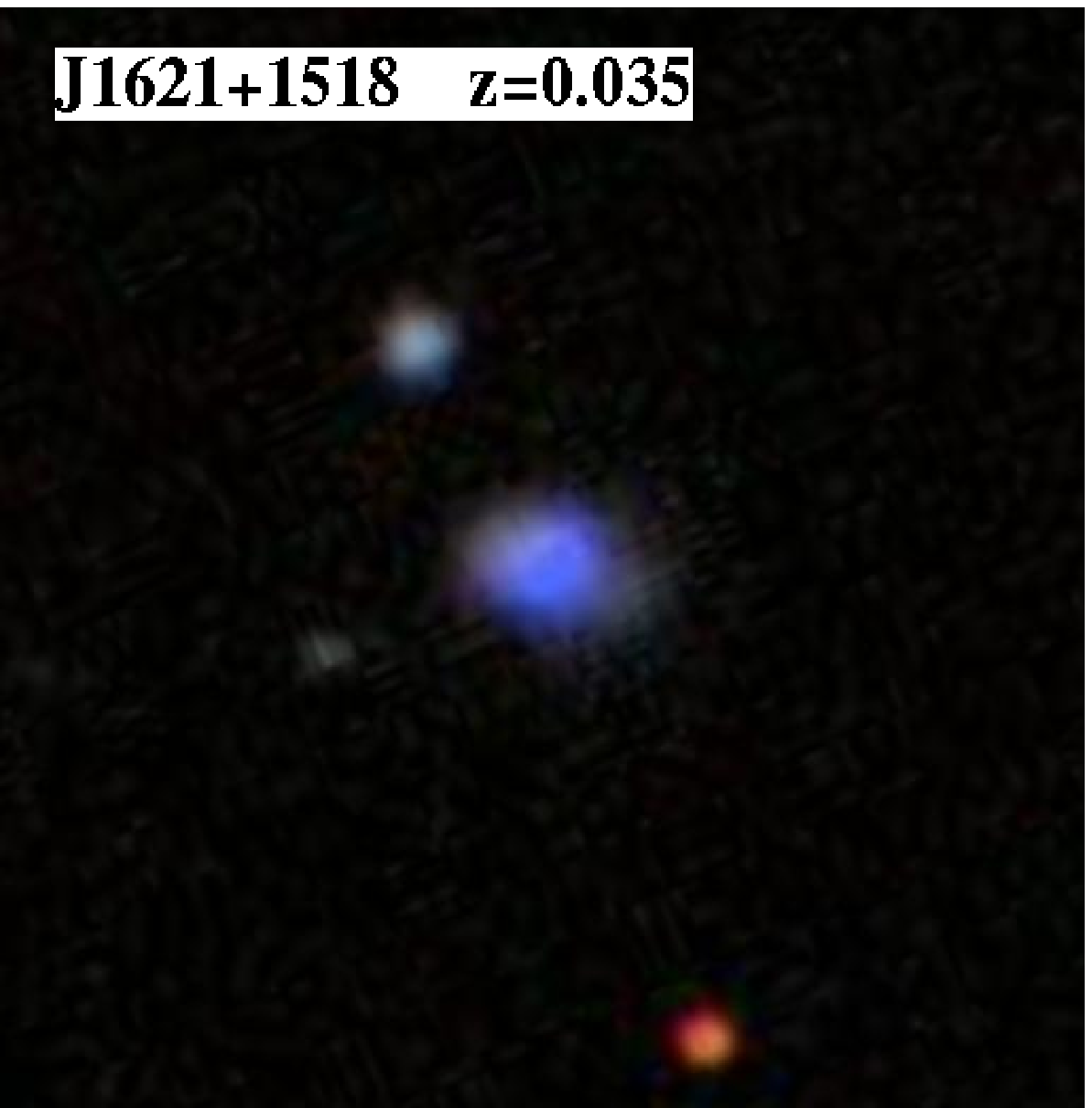} 
\includegraphics[angle=0,width=0.25\linewidth]{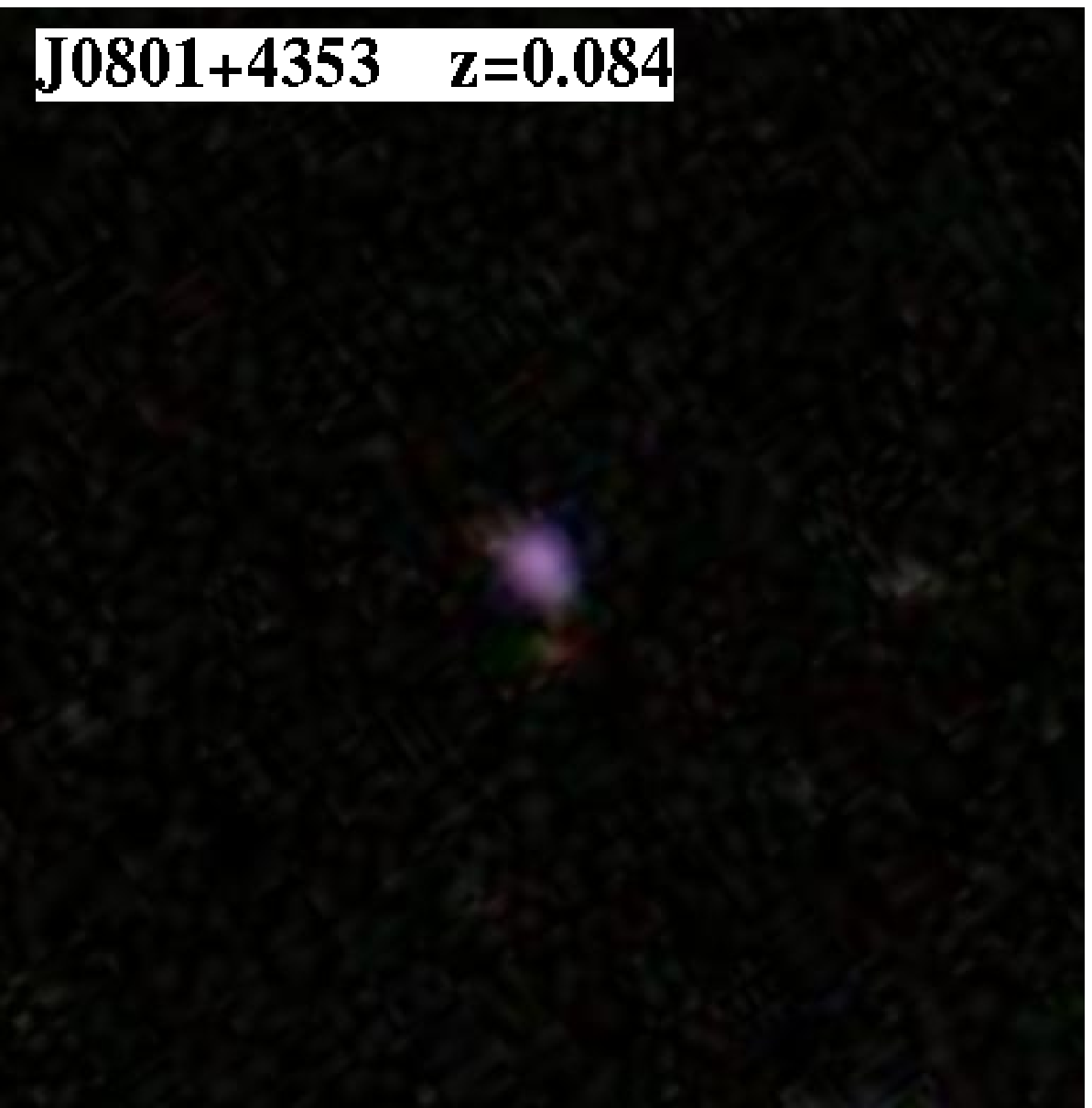} 
\includegraphics[angle=0,width=0.25\linewidth]{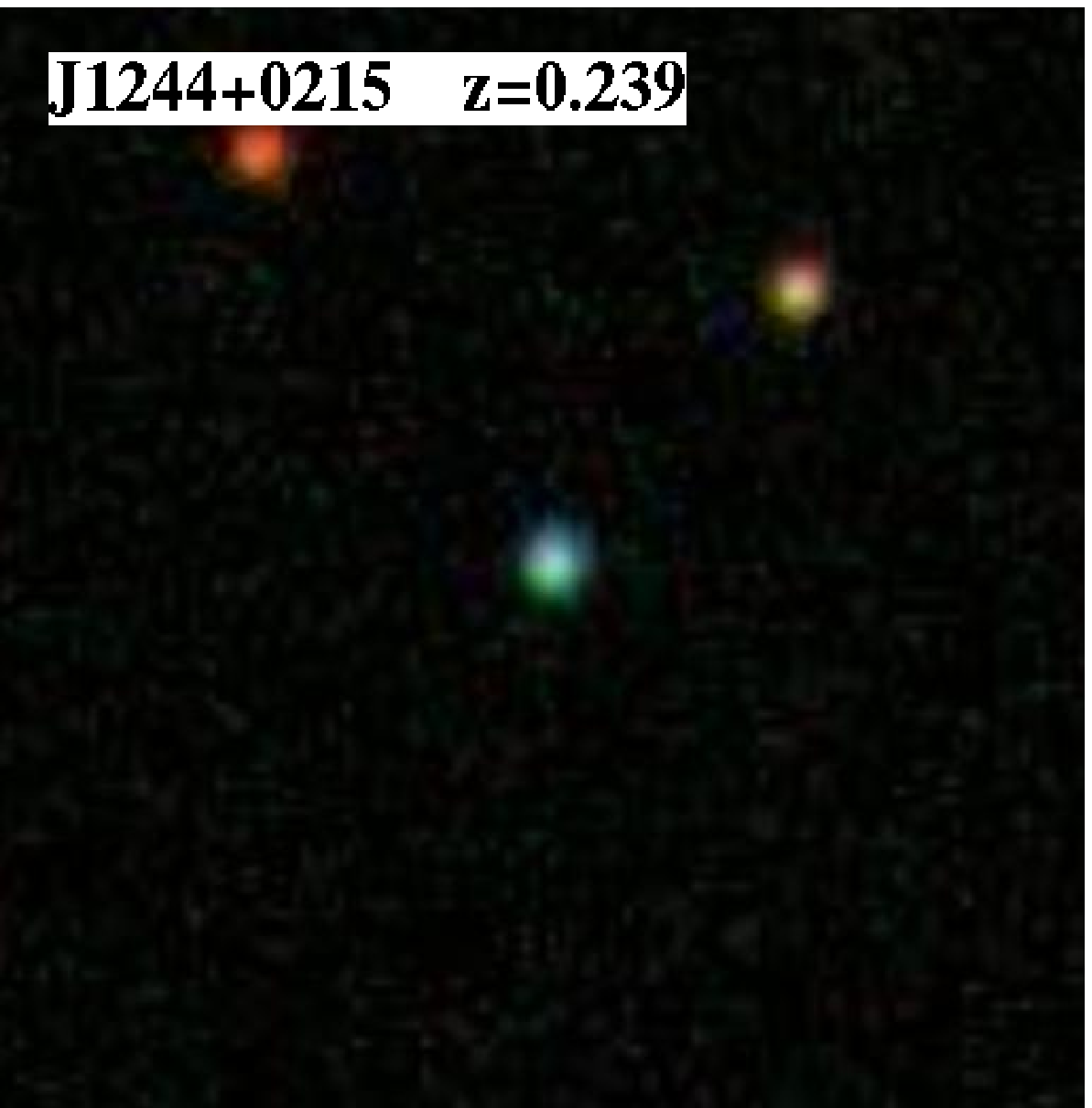} 
\includegraphics[angle=0,width=0.25\linewidth]{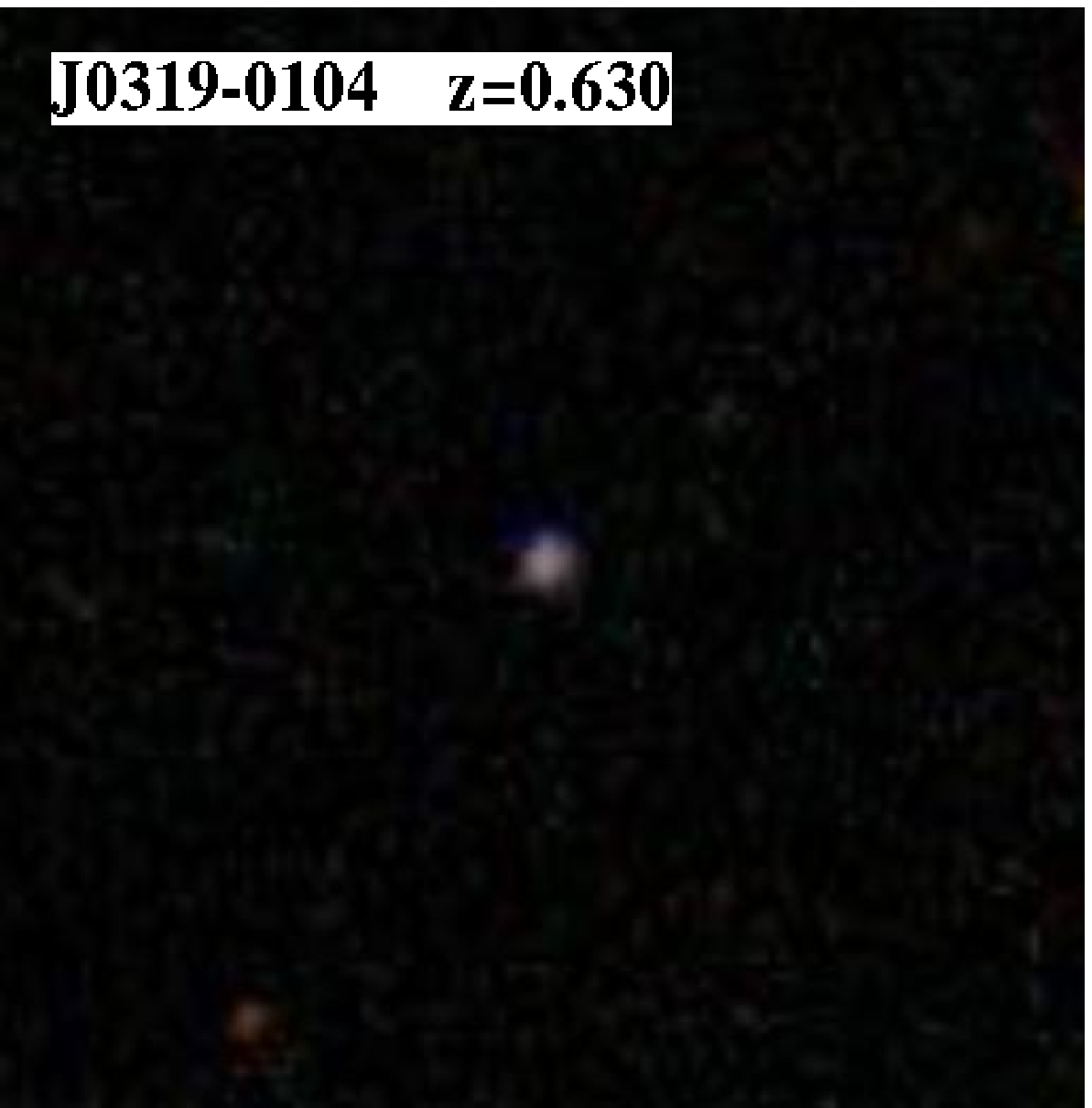}} 
\figcaption{50\arcsec$\times$50\arcsec\ SDSS images of LCGs 
in order of increasing redshift from left to right. 
LCGs from subsample 1 with a round shape are shown in the upper 
panel, while LCGs from subsample 2 with some evidence of 
elongated structure and/or disturbed morphology are shown in the lower panel. 
\label{fig3}}
\end{figure*}

By the same token, green pea galaxies do not 
form a distinct class of objects from LCGs. 
They are just a subsample of the class of LCGs,  
in a restricted redshift range. 
 At other redshifts, the same galaxies would not look green, but blue, 
pink or brown. This is illustrated   
in Fig. \ref{fig3} where we show SDSS images of eight galaxies from our 
LCG sample, located at different distances. 
LCGs in the upper panel are galaxies from subsample 1 with round shape, 
while those  
in the lower panel are galaxies from subsample 2 with extended emission 
and/or  
disturbed morphology. It is seen that, as the strong emission lines (the 
Balmer lines and the [O {\sc iii}] $\lambda$4959\AA, $\lambda$5007\AA\ lines) 
shift through the different SDSS filters, the colors of the LCGs as seen in 
the composite SDSS $gri$ images  
change with redshift. They are blue at low redshifts, becoming
pink at $z$ $\la$ 0.1, green at $z$ $\sim$ 0.2 and brown at
$z$ $\sim$ 0.6. 

\begin{figure*}
\figurenum{4}
\hbox{\includegraphics[angle=-90,width=1.0\linewidth]{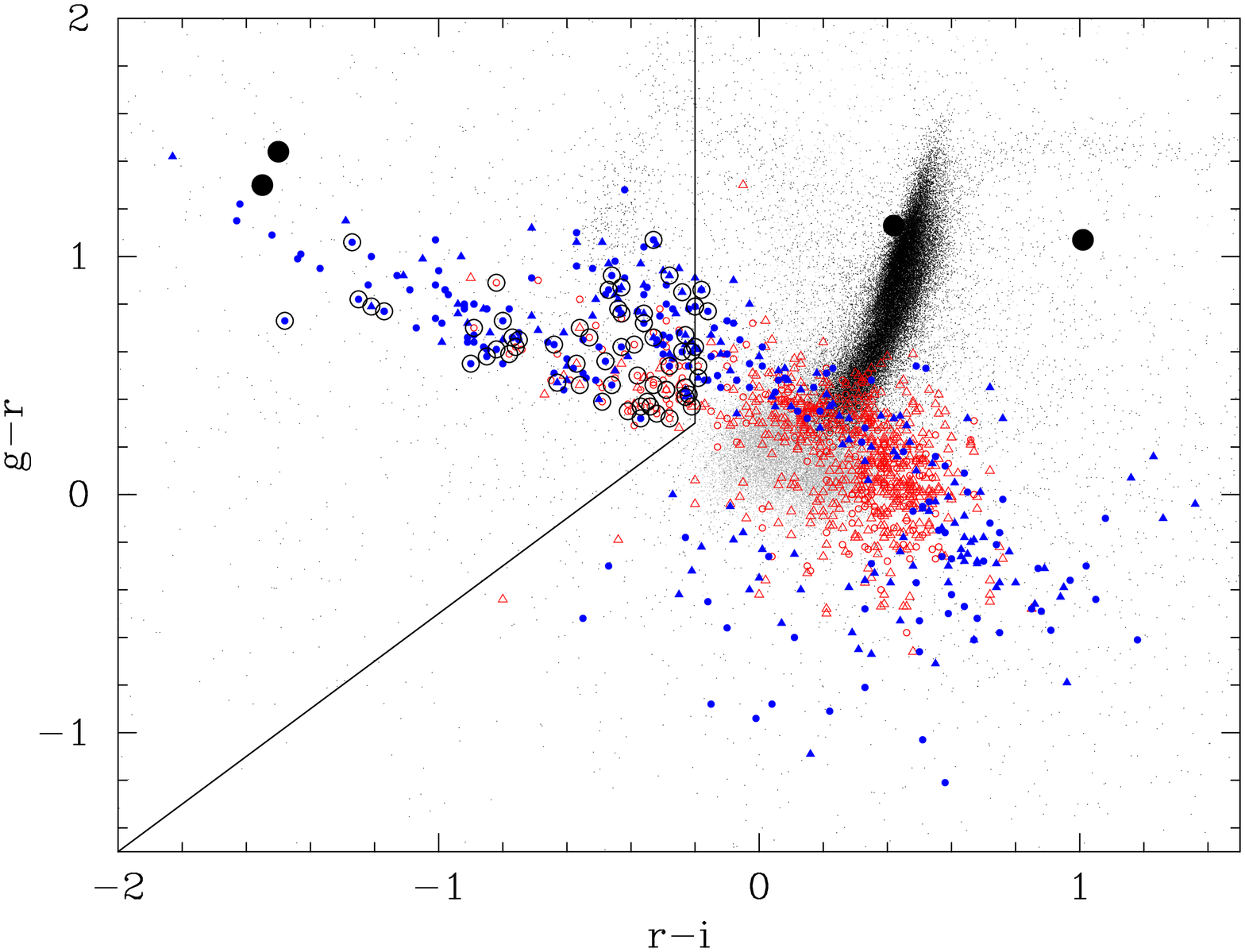}}
\figcaption{$(g-r)$ - $(r-i)$ color-color diagram for LCGs. The symbols 
for the various types of objects are the same 
as in Fig. \ref{fig2}. Galaxies common with the green pea
sample of \citet{C09} are encircled. Low-metallicity AGNs from
\citet{IT08} are shown by large filled circles. The two solid lines are
the dividing lines used by \citet{C09} for selection of 
green pea galaxies. For comparison, a representative sample of
SDSS ``normal'' galaxies (black dots) and 
QSOs \citep{S10} (grey dots) is also shown. 
\label{fig4}}
\end{figure*}

We next compare the colors of the LCGs to those of normal galaxies 
and QSOs.  
We show in Fig. \ref{fig4} the $g-r$ vs $r-i$ color-color diagram for
LCGs (filled and open circles and triangles). For comparison are also shown
representative objects classified in the SDSS as galaxies (black
dots) and QSOs \citep{S10} (grey dots). 
The $g-r$ and $r-i$
colors for LCGs span very large ranges, extending over 
2.5 mag and 3.5 mag respectively. These ranges are the largest for LCGs 
with a high H$\beta$
equivalent width EW(H$\beta$) $\ga$ 100\AA\ (filled symbols). Such large
color ranges are not due to 
differences in galaxy physical properties, but to a large redshift range.
Galaxies with strong emission lines
at $z$ $\sim$ 0.1 -- 0.3 would appear in the left upper corner of 
the diagram, while the same galaxies at $z$ $\sim$ 0.4 -- 0.5 would 
occupy the region with $r-i$ $>$ 1.0, 
in the extreme right part of the diagram. 
The location of the LCGs in 
the $g-r$ vs $r-i$ diagram 
is very different from that of the bulk of galaxies without 
or with weak line emission in their spectra (grey dots). It is interesting 
to note that many LCGs are located in the region with ($g-r$, $r-i$) 
$\sim$ (0.0 -- 0.5, 0.0 -- 0.5) 
populated by QSOs. A significant fraction of these LCGs are automatically
classified in the SDSS as QSOs because of their compact morphology, while
their spectra are that of star-forming galaxies rather than AGNs.

What is the overlap between the green pea sample of \citet{C09} 
and our LCG sample?   
\citet{C09} selected star-forming green pea galaxies 
by requiring them 
to be in the left upper region of the ($g-r$, $r-i$) diagram,  
delineated by the two straight lines in Fig. \ref{fig4}.
We have encircled in this figure all the 66 LCGs that are 
also in the green pea sample of \citet{C09}. The latter sample 
contains a total of 80 objects. 
Out of the 14 missing objects, 12 were 
rejected because of their noisy spectra with a non-detected or weak 
[O {\sc iii}] $\lambda$4363 emission line, although they were present in the 
initial sample. Thus, our LCG 
selection criteria missed only two green pea star-forming 
galaxies out of a total 
of 80 galaxies, i.e. the overlap factor between the green pea and LCG 
samples is 98\%!  
On the other hand, more than half of the LCGs in the region
of green pea star-forming galaxies delineated by the 
two straight lines were not selected by \citet{C09}. All
these galaxies have a green color on the SDSS images and are at redshifts
$z$ $\sim$ 0.1 -- 0.3. 
As for the remaining LCGs, to the right of the 
solid lines in the color-color diagram, they have redshifts $z$ $<$ 0.1 or
$z$ $>$ 0.3, and would not be picked up by the green pea selection criteria, 
although they are the same type of objects, but at different redshifts. 

\subsection{Low-metallicity AGNs in Green Pea Galaxies}

It is interesting to note that the H$\beta$ luminosities of 
the five known low-metallicity compact galaxies that are thought to contain 
AGNs, as evidenced by the strong broad components of their 
H$\alpha$ and H$\beta$ emission lines \citep{IT08,I10} 
put them in the class of LCGs. 
These objects fulfill all the selection criteria discussed above 
for LCGs. They are shown by the large filled circles in Fig. \ref{fig2}.
Their redshifts are in the range 0.1--0.3, 
and therefore they can also be classified as green pea galaxies. However, 
in the ($g-r$) - ($r-i$) color-color diagram in Fig. \ref{fig4}, 
only two low-metallicity AGNs fall 
into the green pea region. The two other low-metallicity AGNs 
are located outside both the 
green pea and LCG regions. The latter two objects 
show abnormally red $g-r$ and $r-i$ colors, 
probably due to high dust extinction. The fifth low-metallicity AGN
studied by \citet{I10} is not shown in Fig. \ref{fig4} because it is beyond
the SDSS sky area and therefore we have no $g,r,i$ colors for it.
 
It is not unreasonable that AGNs tend to occur in more luminous and 
hence more massive compact galaxies, as these have more massive star formation 
and hence are more favorable to the formation of intermediate-mass black holes.
 It would be interesting to obtain 
high-resolution and high signal-to-noise ratio
spectra of more objects in the LCG sample to look for 
more objects with broad components which may contain AGNs.  

\subsection{The [O {\sc iii}] $\lambda$5007/H$\beta$ vs. [N {\sc ii}] $\lambda$6583/H$\alpha$ diagram}
 
\begin{figure*}
\figurenum{5}
\hbox{\includegraphics[angle=-90,width=1.0\linewidth]{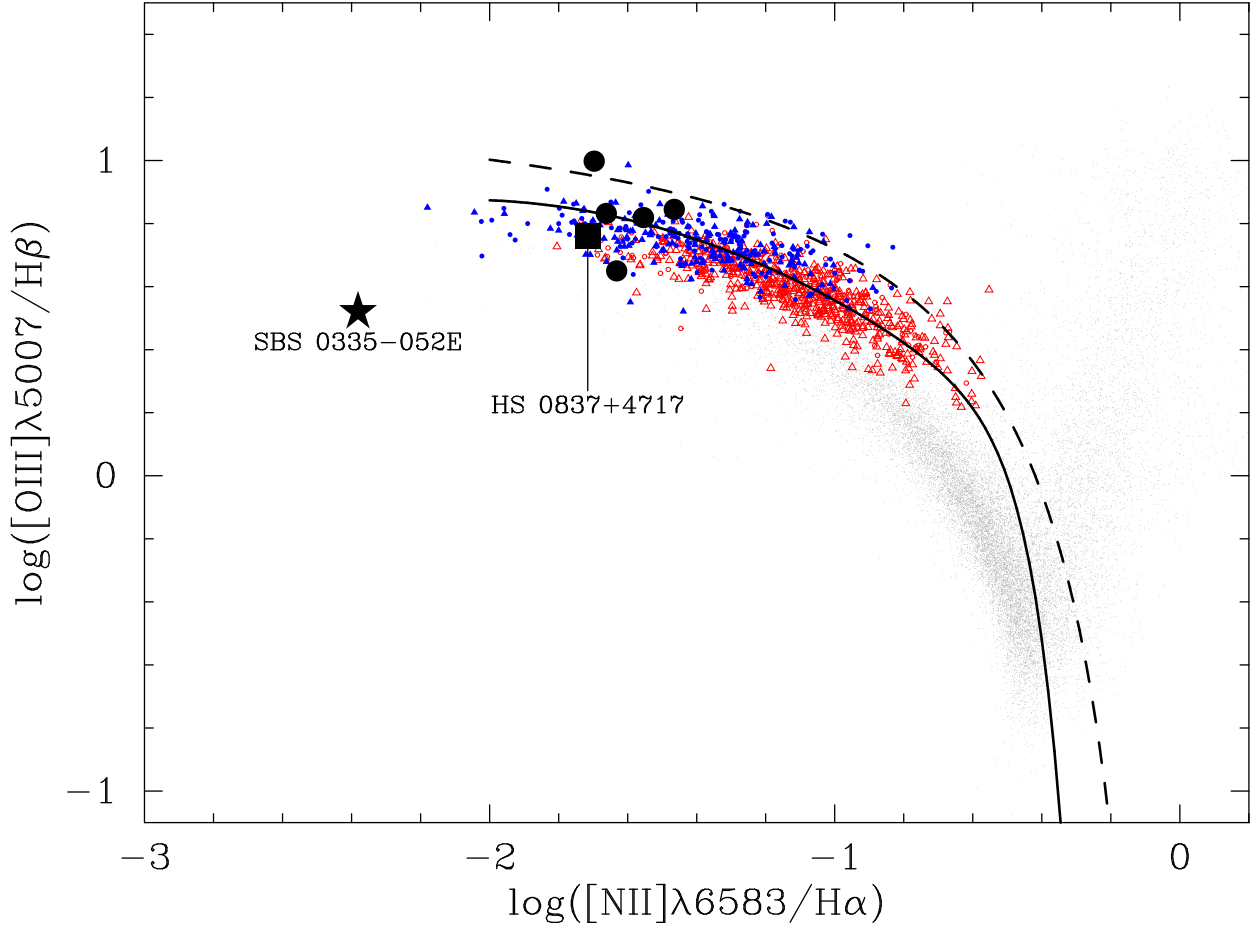}}
\figcaption{The Baldwin-Phillips-Terlevich (BPT) diagram \citep{B81}
for narrow emission lines. LCGs
are shown by small open and filled symbols, 
their meaning being the same as in Fig. \ref{fig2}. 
Also, plotted are the 100 000
emission-line galaxies from SDSS DR7 (cloud of grey dots), the five 
low-metallicity AGNs from 
\citet{IT08} and \citet{I10} (filled circles), and the two well-studied
BCDs SBS 0335--052E and HS 0837+4717. The dashed line
from \citet{K03} and the solid line from \citet{S06} separate star-forming 
galaxies from active galactic nuclei. \label{fig5}}
\end{figure*}

Figure \ref{fig5} shows by small filled and open symbols the positions of LCGs
in the 
[O {\sc iii}] $\lambda$5007/H$\beta$ vs. [N {\sc ii}] $\lambda$6583/H$\alpha$
diagram \citep[][hereafter BPT]{B81}.
 The star and filled square represent respectively 
the well-studied BCDs SBS 0335--052E and HS 0837+4717.
HS 0837+4717  lies in the same region as the LCGs while SBS 0335--052E 
lies 
to the left, having an abnormally low [N {\sc ii}] $\lambda$6583/H$\alpha$
ratio. This is because SBS 0335--052E has a 
low N/O abundance ratio, a characteristic
of extremely low-metallicity BCDs \citep[e.g. ][]{I06}, and the
ionizing parameter of its H {\sc ii} region is high, resulting in a small
abundance of singly ionized nitrogen.
For comparison, the grey dots represent
all galaxies from the SDSS DR7 with flux errors smaller
than 10\% for each of the four emission lines H$\beta$, [O {\sc iii}]
$\lambda$5007, H$\alpha$ and [N {\sc ii}] $\lambda$6583. The emission line 
fluxes for the SDSS galaxies shown by grey dots were measured using the 
technique developed by \citet{T04} and are taken from the SDSS website
\footnote{http://www.sdss.org/DR7/products/value$_{-}$added/index.html.}. These
galaxies are distributed into two wings, the left one thought to contain 
star-forming galaxies and the right one AGN hosts. The dashed line 
represents the empirical divisory line between star-forming galaxies and AGNs 
proposed by \citet{K03}, while the continuous line represents the upper limit 
for pure star-forming galaxies taken from \citet{S06}. 
We see that the LCGs 
lie in the low-metallicity part of what 
is usually considered as the region of star-forming galaxies. 
We note that the five low-metallicity AGN candidates from \citet{IT08} and 
\citet{I10}, represented by large filled circles, lie in the same 
region as the LCGs. The BPT diagram cannot distinguish between 
star-forming galaxies and AGN hosts in the low-metallicity regime. 

\section{Element abundances}

We derived element abundances in all 803 LCGs 
from the narrow emission-line fluxes, using the 
so-called direct method. This method is based on the determination of 
the electron temperature within the [O {\sc iii}] zone 
from the [O {\sc iii}]$\lambda$4363/($\lambda$4959+$\lambda$5007) line ratio.
The fluxes in all spectra were measured using Gaussian fitting with the
IRAF\footnote{IRAF is the Image Reduction and Analysis Facility distributed
by the National Optical Astronomy Observatory, which
is operated by the Association of Universities for Research in Astronomy 
(AURA) under cooperative agreement with the
National Science Foundation (NSF).} SPLOT routine. They were corrected for 
both extinction, using the reddening curve of \citet{W58}, and underlying 
hydrogen stellar absorption, derived simultaneously by an iterative procedure 
described by \citet{I94} and using the observed decrements of the narrow
hydrogen Balmer lines. The extinction coefficient $C$(H$\beta$) and equivalent 
width of hydrogen absorption lines EW(abs) are derived in such a way to obtain 
the closest agreement between the extinction-corrected and theoretical 
recombination hydrogen emission-line fluxes normalized to the H$\beta$ flux. 
It is assumed that EW(abs) is the same for all hydrogen lines. This assumption 
is justified by the evolutionary stellar population synthesis models of 
\citet{G05}.

The physical conditions, and the ionic and total heavy element abundances in 
the H {\sc ii} regions were derived following \citet{I06}. In particular for 
O$^{2+}$, Ne$^{2+}$, and Ar$^{3+}$ abundances, we adopt the temperature 
$T_e$(O {\sc iii}) directly derived from the [O {\sc iii}]
$\lambda$4363/($\lambda$4959 + $\lambda$5007) emission-line ratio. The 
electron temperatures $T_e$(O {\sc ii}) and $T_e$(S {\sc iii}) were derived 
from the empirical relations by \citet{I06}. We used $T_e$(O {\sc ii})
for the calculation of O$^+$, N$^+$, S$^+$, and Fe$^{2+}$ abundances
and $T_e$(S {\sc iii}) for the calculation of S$^{2+}$, Cl$^{2+}$, and 
Ar$^{2+}$ abundances. The electron number densities $N_e$(S {\sc ii}) were 
obtained from the [S {\sc ii}] $\lambda$6717/$\lambda$6731 emission-line 
ratios, respectively. The low-density limit holds for the H {\sc ii} regions 
that exhibit the emission lines considered here. The element abundances then 
do not depend sensitively on $N_e$. We use the ionization correction factors
(ICFs) from \citet{I06} to correct for unseen stages of ionization and
to derive the total O, N, Ne, S, Cl, Ar, and Fe abundances.

\begin{figure*}
\figurenum{6}
\hbox{\includegraphics[angle=-90,width=0.49\linewidth]{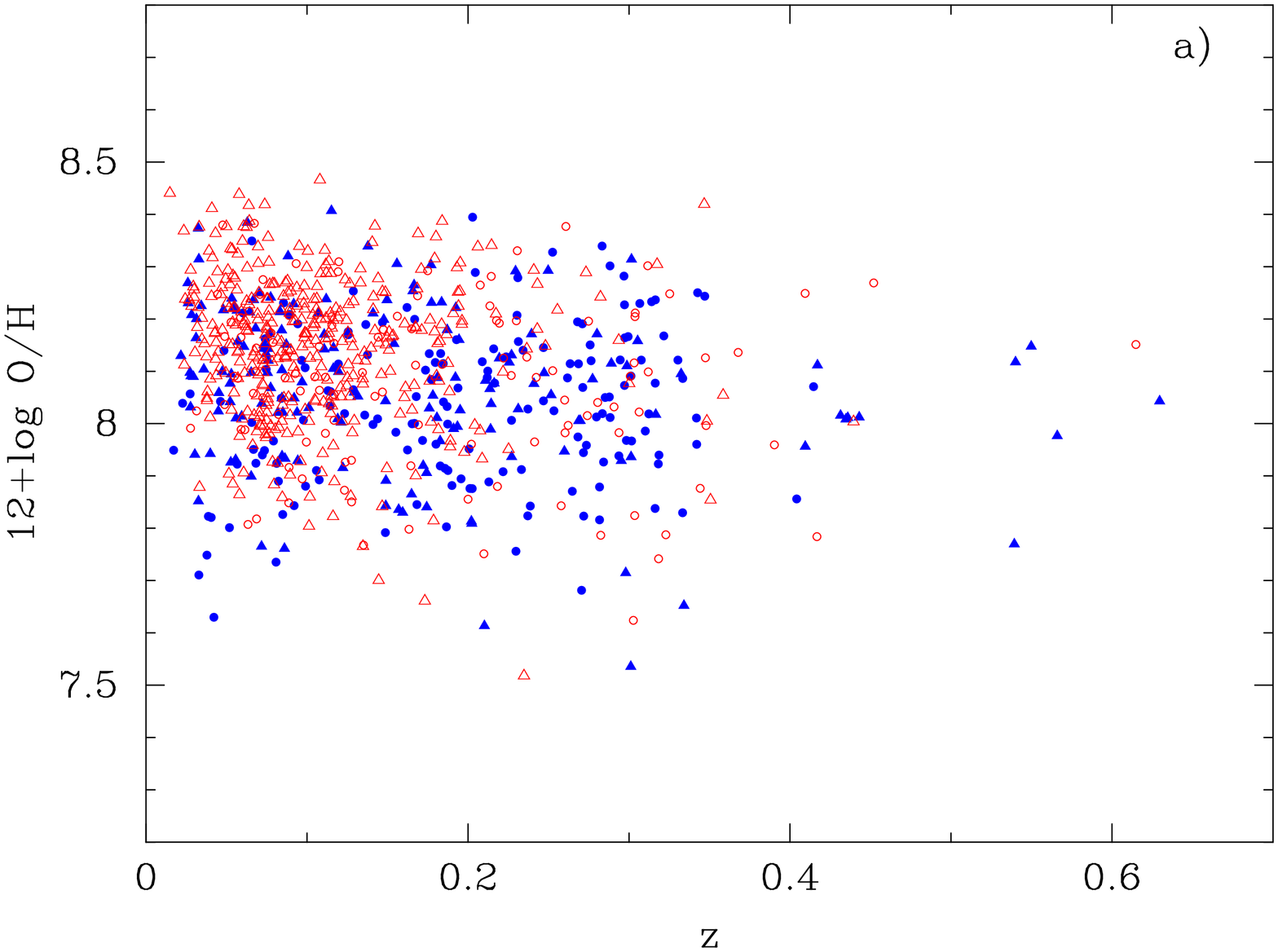} 
\includegraphics[angle=-90,width=0.49\linewidth]{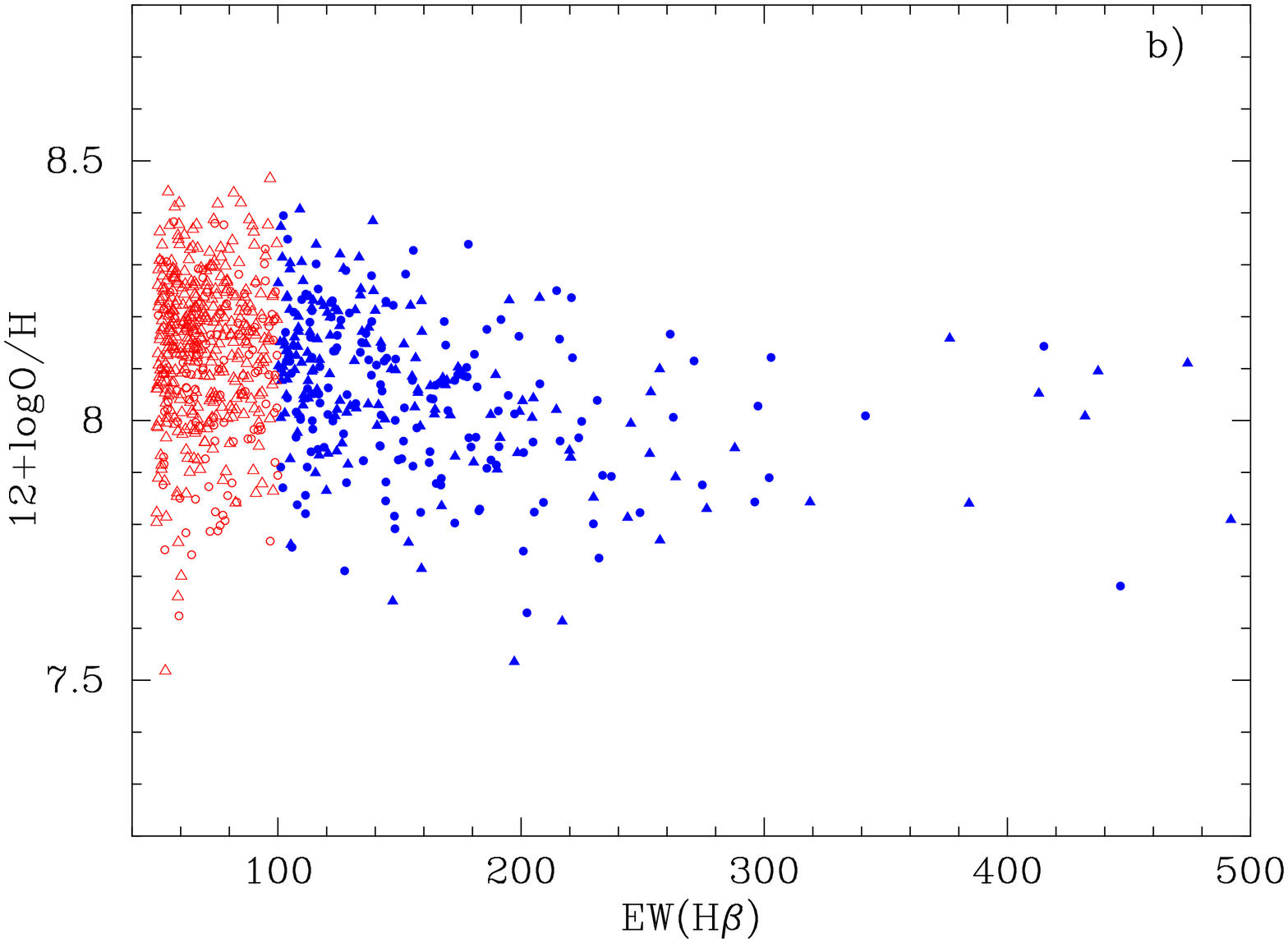}}
\figcaption{ Oxygen abundance 12 + log O/H  
(a) vs. redshift $z$ and (b) vs. the equivalent width EW(H$\beta$) of the 
H$\beta$ emission line. The LCGs are shown by symbols with 
the same meaning as in Fig. \ref{fig2}.
\label{fig6}}
\end{figure*}

The dependence of the oxygen abundance 12 +log O/H of the 
LCGs on their redshift $z$
is shown in Fig. \ref{fig6}a. There is no clear variation of 12+log O/H 
with $z$. The median value of 12 + log O/H for the whole sample 
is $\sim$ 8.11 or 0.26 solar (Table \ref{tab1}).  
This median value is similar to the one obtained by \citet{A10} for the
subsample of the green pea galaxies of \citet{C09} they used. 
However, our values
of 12 +log O/H are significantly lower -- by $\sim$ 0.6 dex -- 
than those obtained
by \citet{C09}. This difference is due to the 
different techniques used to derive abundances.
We have used the direct method while \citet{C09}
have used an empirical calibration method based on the 
[N {\sc ii}] $\lambda$6583/H$\alpha$ flux ratio, despite the fact that nearly
all their spectra contain a strong enough [O {\sc iii}] $\lambda$4363 emission
line, which can be used to 
determine the electron temperature of the [O {\sc iii}] zone.  
 Evidently, there is an offset between oxygen 
abundances derived by the direct method and 
those derived by the strong-line empirical method used by \citet{C09}. 
A more detailed comparison between abundances obtained by us and by 
\citet{C09} is given in Section 5.3.

Fig. \ref{fig6}a also shows 
that LCGs with lower equivalent widths EW(H$\beta$) 
(open symbols)  
have systematically higher 12 + log O/H than 
LCGs with higher equivalent widths (filled symbols). 
While the median 12 + log O/H is 8.04 for  
subsample 1, it is 8.16 for subsample 4 (Table \ref{tab1}).
This effect is more pronounced in Fig. \ref{fig6}b where we show the dependence
of 12 + log O/H on EW(H$\beta$). 
The LCGs with high EW(H$\beta$) $>$ 200\AA\ 
are more metal-poor than LCGs with lower EW(H$\beta$) by $\sim$ 0.2 dex.
Such variations can be explained by the fact that LCGs with a lower 
EW(H$\beta$) are in general more massive and thus more metal-rich
(see the discussion of the mass-metallicity relation below).

\begin{figure*}
\figurenum{7}
\hbox{\includegraphics[angle=0,width=0.9\linewidth]{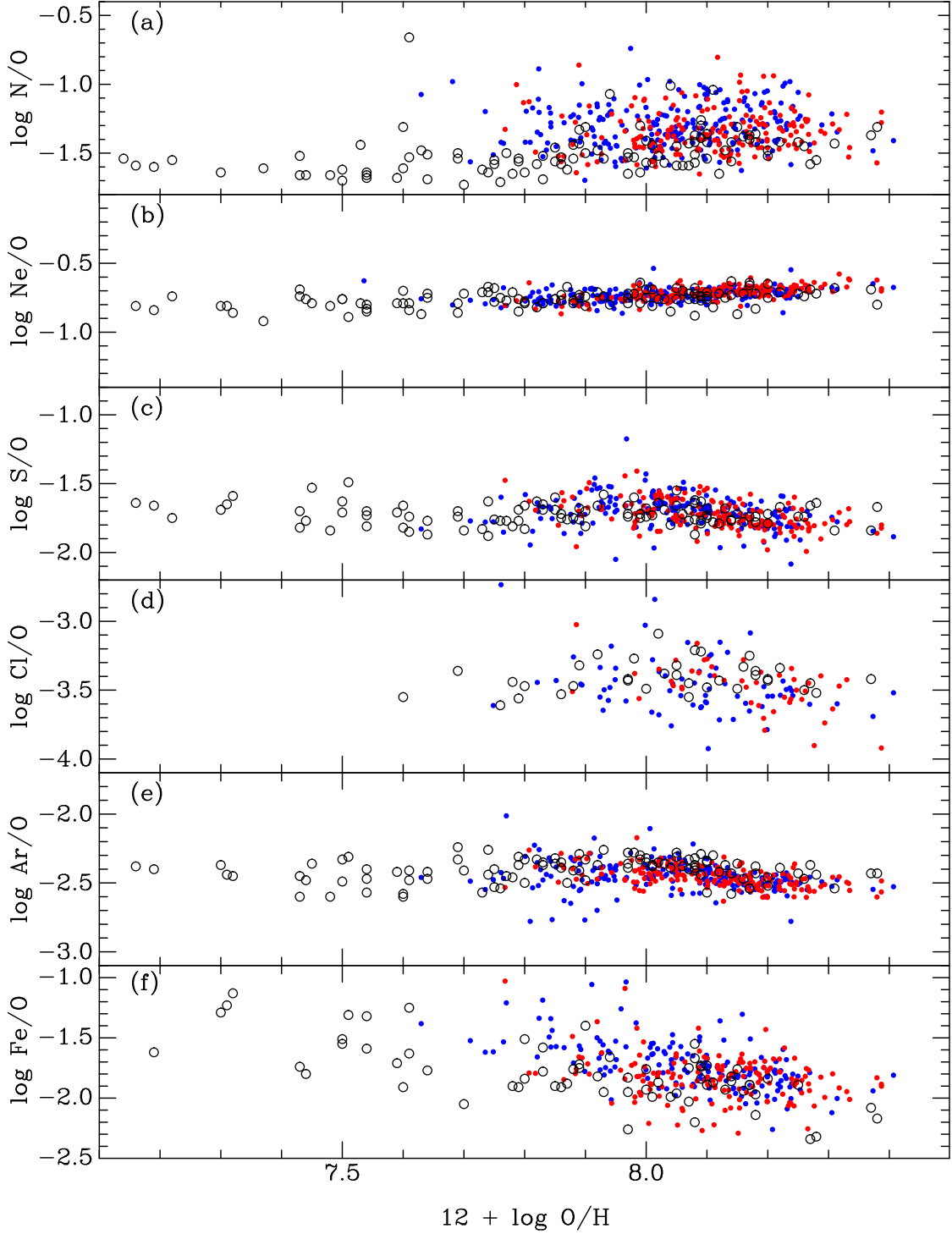}} 
\figcaption{Abundance ratios (a) log N/O, (b) log Ne/O, (c) log S/O,
(d) log Cl/O, (e) log Ar/O and (f) log Fe/O vs. oxygen
abundance 12 + log O/H for emission-line galaxies. Large open
circles denote galaxies from the HeBCD sample \citep{I94,IT04}, 
and small filled circles represent LCGs (this paper).
\label{fig7}}
\end{figure*}

In Fig. \ref{fig7}, we show the dependence of the 
N/O, Ne/O, S/O, Cl/O, Ar/O and
Fe/O abundance ratios on the oxygen abundance 12 +log O/H for the 
LCG sample (filled circles).
For comparison, we also show a sample of BCDs
that has been collected primarily to study helium abundances in
low-metallicity dwarf galaxies.
This sample is the same as in \citet{I94}, with the addition of galaxies from 
\citet{IT04} and includes mainly nearby low-metallicity and low-luminosity
galaxies.
It is seen from Fig. \ref{fig7} that the distribution of 
heavy element abundance ratios in LCGs
is very similar to those in BCD galaxies, 
despite the fact that LCGs have higher oxygen abundances 
(12 + log O/H $\ga$ 7.6), while the BCD galaxies span a larger range of 
metallicities, with 12 + log O/H as low as 7.12. This implies that 
the chemical enrichment history has been very similar in BCDs and LCGs.

The N/O vs O/H diagram (Fig. \ref{fig7}a) shows a significant spread of N/O
values, in the range --0.9 - --1.7 at 12 + log O/H $\ga$ 7.6,
 for both the LCG and 
BCD samples. However, there is no appreciable increase of N/O with 
increasing 12 + log O/H up to $\sim$ 8.4, suggesting that nitrogen is mainly of primary origin in LCGs. 
This is at variance with the results of \citet{A10} who find a clear
increase of N/O with 12 + log O/H over the whole 7.5 - 8.4 range, suggesting that secondary production of nitrogen is important. Additionally, \citet{A10} 
derived a high log N/O, larger than the solar ratio of --0.86 \citep{As09}, 
in a significant number of
green pea galaxies with 12 + log O/H in the range $\sim$ 8.1 - 8.4. 
The reason for the differences between our results and those of \citet{A10} is not clear. We note that \citet{A10} used 
the pipeline measurements of line fluxes available in the SDSS archive, 
while we 
measured them directly using the SDSS spectra, taking into account the possible contamination of [N {\sc ii}] $\lambda$6583 by H$\alpha$ $\lambda$6563. 

It is worth noting the increase of the Ne/O abundance ratio and the decrease of the Fe/O ratio with
increasing oxygen abundance. These trends are caused by the depletion of O and
Fe onto dust grains, as discussed by \citet{I06}. The LCG data 
confirm these trends.

\section{Luminosity-metallicity relation}

\begin{figure*}
\figurenum{8}
\hbox{\includegraphics[angle=-90,width=1.0\linewidth]{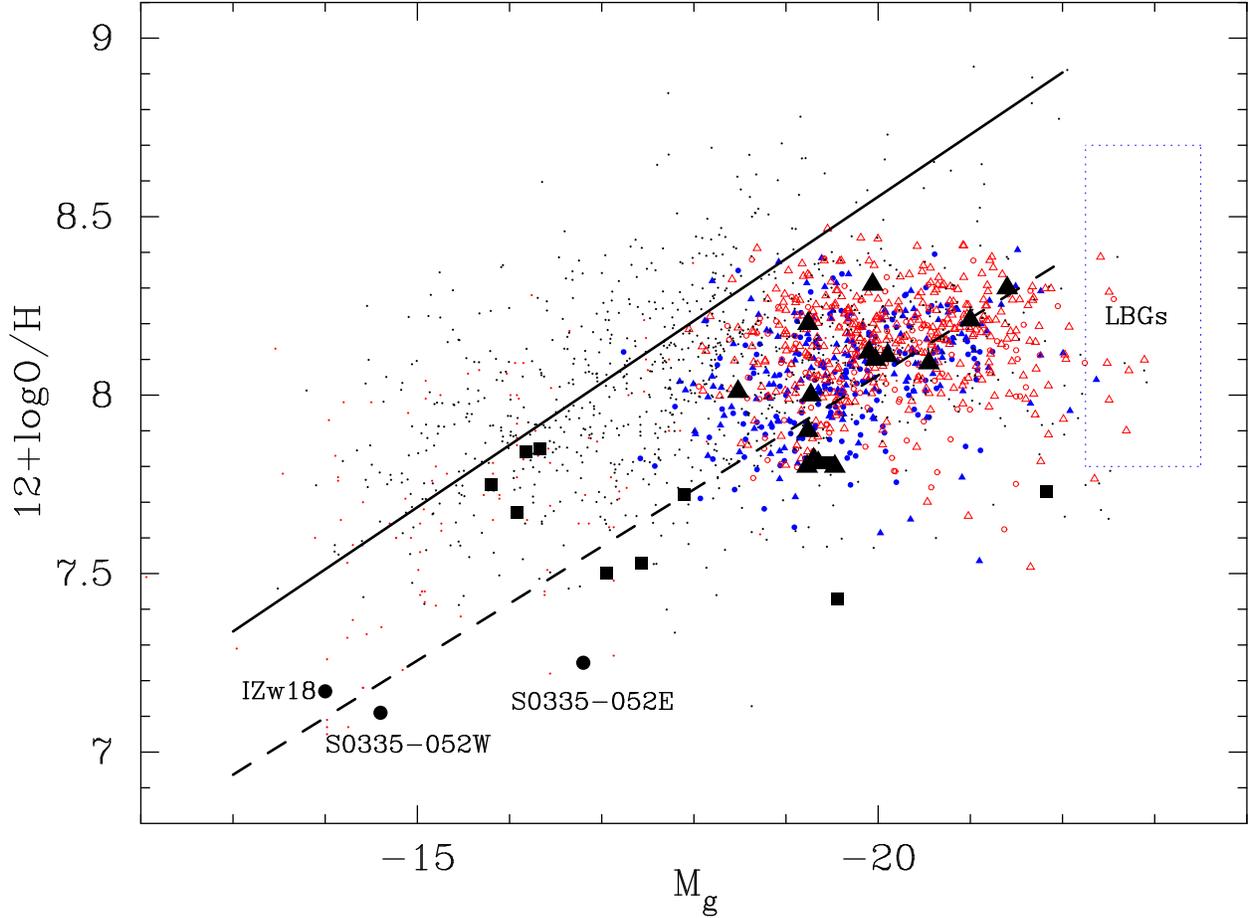}} 
\figcaption{Luminosity-metallicity relation for emission-line galaxies.
LCGs are shown by the same symbols as in Fig. \ref{fig2} (filled and open
circles and triangles). The emission-line galaxies studied by 
\cite{G09} are shown by dots. The luminosity-metallicity relation for these
galaxies is fitted by a solid line \citep{G09}. The most metal-poor local BCDs
-- I Zw 18, SBS 0335--052W and SBS 0335--052E-- are shown by large filled
circles \citep{G09}. The extremely metal-poor emission-line galaxies at 
intermediate redshifts $z$ $<$ 1 studied by \citet{K07} and the 
luminous metal-poor
star-forming galaxies at $z$ $\sim$ 0.7 studied by \citet{H05} 
are shown by large filled squares
and large filled triangles, respectively. The region occupied by
Lyman-break galaxies (LBGs) at $z$ $\sim$ 3 studied 
by \citet{P01} is indicated by 
a dotted rectangle. The linear best likelihood fit to the strongly 
star-forming galaxies, excluding LBGs, is shown by a dashed line.
\label{fig8}}
\end{figure*}

The relation between the oxygen abundance 12 + log O/H and the extinction-corrected
absolute magnitude $M_g$ in the SDSS $g$ band for LCGs is shown in 
Fig. \ref{fig8}.  
The symbols for LCGs are the same as in Fig. \ref{fig2}.
The $g$ apparent magnitudes are taken from the SDSS data base. The correction
for extinction is done using the extinction coefficient $C$(H$\beta$) obtained
from the hydrogen Balmer decrement in the SDSS optical spectra. Distances of
LCGs are derived from redshifts and adopting a Hubble constant 
$H_0$ = 75 km s$^{-1}$ Mpc$^{-1}$. For comparison, we show by dots the 
emission-line galaxy sample of \citet{G09}. 
We also show by large filled circles the three most
metal-deficient BCDs known in the local universe, I Zw 18, SBS 0335--052W
and SBS 0335-052E \citep{G09}. The intermediate-redshift ($z$ $<$ 1) extremely
low-metallicity emission-line galaxies studied by \citet{K07} 
are shown by large
filled squares, while the luminous metal-poor star-forming galaxies at 
$z$ $\sim$ 0.7 studied by \citet{H05} are shown by large filled triangles.
The dotted-line rectangle indicates the location of the Lyman-break galaxies 
(LBG) at $z$ $\sim$ 3 of \citet{P01}. The solid line is the luminosity-metallicity relation derived by \citet{G09} for their emission-line galaxy sample (solid line in their Fig. 9).

LCGs are clearly offset by $\sim$ 2 mag to brighter magnitudes as compared to
the bulk of the local emission-line galaxies, implying strong ongoing star
formation. There exists a similar offset for the three most-metal
deficient BCDs I Zw 18, SBS 0335--052W and SBS 0335--052E,  
the intermediate-redshift extremely metal-poor emission-line galaxies of 
\citet{K07}, the luminous metal-poor emission-line galaxies at $z$ $\sim$ 0.7
of \citet{H05} and the LBGs of \citet{P01}. It appears that all these types of galaxies are undergoing strong bursts of star formation and that they 
define a common luminosity-metallicity relation, the one for actively star-forming 
galaxies, going from the most-metal deficient BCDs at the faint end to LBGs at 
the bright end. The best linear likelihood fit to the strongly star-forming 
galaxies, excluding LBGs for which the oxygen
abundances are derived by using not the direct but the strong-line method, is 
\begin{equation}
12+\log {\rm O/H} = (-0.16 \pm 0.01)\times M_g + (4.86\pm 0.19).
\end{equation}
It is shown by a dashed line in Fig. \ref{fig8}. 

Our LCGs, despite
their lower redshift range ($z$ $\sim$ 0.02 -- 0.63),
 are very similar in luminosity to the luminous
galaxies at $z$ $\sim$ 0.7 of \citet{H05}, but they are, in the mean, some  3 mag fainter than LBGs.
However, the most luminous objects in our LCG sample  
do have oxygen abundances and luminosities that are  
similar to those of LBGs at $z$ $\sim$ 3, 
implying that these two types of objects are likely
of the same nature. In particular, most of the LCGs inside the LBG dotted 
rectangle (Fig. \ref{fig8}) have disturbed
morphologies and relatively low equivalent widths EW(H$\beta$). 
This implies some kind of gravitational interactions and/or merging processes 
in the galaxies with the significant fraction of
a non-ionizing stellar population that contributes non-negligibly to the observed stellar  
continuum. The similarity in properties between LCGs and LBGs suggests
 that, since LCGs are 
relatively nearby, the bulk of them being in the redshift range 0 -- 0.3, 
they can be studied in detail to shed light 
on the physical properties of distant LBGs with $z$ $\sim$ 3.

\section{Stellar masses}

The stellar mass of a galaxy is one of its most important global
characteristics. It can be derived by modeling its spectral energy
distribution (SED) which depends on the adopted
star formation history. However, in the case of LCGs, the situation
is complicated by the presence of strong ionized gas emission, which must 
be subtracted before stellar mass determination. \citet{C09} did consider
the effect of the strong nebular emission lines. However, they did not 
take into account the effect of the gaseous continuum emission which is
significant in spectra of LCGs with a high EW(H$\beta$). 
Neglecting the correction
for gaseous continuum emission would result in a significant overestimate of
the galaxy stellar mass for two reasons: 
1) gaseous continuum emission increases the luminosity of the galaxy; 
and 2) the SED of gaseous continuum emission is flatter than that of stars,   
making the SED redder than expected
for pure stellar emission. Consequently, the fraction of the light
from the red old stellar population is artificially increased.
Therefore, to derive the correct stellar mass of the galaxy, we have used
the method described below that takes into account the contribution of gaseous continuum emission.

\subsection{Method}

The method consists of fitting a series of model SEDs to the observed one and finding the best
fit. It is described in \citet{G06,G07} and consists of the following.
The fit is performed for each SDSS spectrum over the whole
observed spectral range of $\lambda$$\lambda$3900--9200\AA, which 
includes the Balmer jump region ($\lambda$3646\AA) for high-redshift LCGs 
with $z$ $>$ 0.1, and the Paschen jump region ($\lambda$8207\AA) for
LGCs with $z$ $<$ 0.12. 
As each SED is the sum of both stellar and ionized gas emission,
its shape depends on the relative contribution of these two
components. In LCGs with high EW(H$\beta$) $>$ 100\AA\ (subsamples 1 and 2), the 
contribution of the ionized gas emission can be very large. However, the 
equivalent widths of the hydrogen emission lines never reach the theoretical 
values for pure gaseous emission, which for the H$\beta$ emission line
is $\sim$ 900 -- 1000\AA, slightly depending on the electron temperature 
of the ionized gas. The objects in our LCG sample span a lower range of EW(H$\beta$)s, 
between 50 and $\sim$500\AA\ (Fig. \ref{fig6}b), due to the
contribution of stellar continuum emission. 
The contribution of gaseous 
emission relative to stellar emission can be parameterized by the equivalent
width of the H$\beta$ emission line EW(H$\beta$). Given a temperature
$T_e$(H$^+$), the ratio of the gaseous emission to the total emission is equal 
to the ratio of the observed EW(H$\beta$) to the equivalent width of H$\beta$
expected for pure gaseous emission. The shape of the spectrum
depends also on reddening. The extinction coefficient for the ionized
gas $C$(H$\beta$)$_{\rm gas}$ can be obtained from the observed hydrogen
Balmer decrement. However, there is no direct way to derive the
extinction coefficient $C$(H$\beta$)$_{\rm stars}$ for the stellar emission, 
which can be in principle different from the ionized gas extinction coefficient.
Finally, the SED depends on the star formation history of the galaxy.

We have carried out a series of Monte Carlo simulations to
reproduce the SED of each galaxy in our sample. To calculate
the contribution of the stellar emission to the SEDs, we have
adopted the grid of the Padua stellar evolution models by \citet{Gi00}
with heavy element mass fractions $Z$ = 0.001, 0.004, and 0.008. To reproduce 
the SED of the stellar component with any star formation history, we have 
calculated with the package PEGASE.2 \citep{FR97} a grid
of instantaneous burst SEDs for a stellar mass of 1 $M_\odot$ in a wide range 
of ages, from 0.5 Myr to 15 Gyr. We have adopted a stellar initial mass 
function with a Salpeter slope, an upper mass limit of 100 $M_\odot$, and a 
lower mass limit of 0.1 $M_\odot$. Then the SED with any star formation 
history can be obtained by integrating the instantaneous burst SEDs over
time with a specified time-varying star formation rate. 

We have approximated the star formation history in each LCG by a recent short 
burst with age $t$(young) $<$ 10 Myr, which accounts 
for the young stellar population, and a prior continuous star formation responsible for 
the older stars, with 
age starting at $t_2$ $\equiv$ $t$(old) and ending at $t_1$, with 
$t_2$ $>$ $t_1$ and varying between 10 Myr and 15 Gyr.
The contribution of each stellar population to the SED is parameterized by the ratio
of the masses of the old to young stellar populations, $b$ = $M$(old)/$M$(young), which we vary between 0.01 and
1000. 

The contribution of the gaseous emission was scaled to the
stellar emission using the ratio of the observed equivalent width
of the H$\beta$ emission line to the equivalent width of H$\beta$ expected
for pure gaseous emission. Then the total modeled monochromatic
(gaseous+stellar) continuum flux near the H$\beta$ emission line for a 
mass of 1 $M_\odot$ is scaled to fit the monochromatic 
extinction-corrected luminosity of 
the galaxy at the same wavelength. The scaling factor is equal to the total
stellar mass $M_*$ in solar units. In our fitting model 
$M_*$= $M$(young) + $M$(old), where $M$(young) and $M$(old) are respectively the masses
of the young and old stellar populations in solar units. 
These masses are derived using $M_*$ and $b$.

The SED of the gaseous continuum is
taken from \citet{A84} and includes hydrogen and helium free-bound,
free-free, and two-photon emission. In our models, it is
always calculated with the electron temperature $T_e$(H$^+$) of the
H$^+$ zone and with the chemical composition derived from the
H {\sc ii} region spectrum. The observed emission lines corrected for
reddening and scaled using the flux of the H$\beta$ emission
line were added to the calculated gaseous continuum. 
The flux ratio of the gaseous continuum to the total continuum depends
on the adopted electron temperature $T_e$(H$^+$) in the H$^+$ zone,
since EW(H$\beta$) for pure gaseous emission decreases with increasing
$T_e$(H$^+$). Given that $T_e$(H$^+$) is not necessarily equal to
$T_e$(O {\sc iii}), we chose to vary it in the range 
(0.7-1.3)$\times$$T_e$(O {\sc iii}).
Strong emission lines in LCGs (see Fig. \ref{fig1}) are measured
with good accuracy, so the equivalent width of the H$\beta$ emission
line and the extinction coefficient for the ionized gas are accurate
to 5\% and 20\%, respectively. Thus, we vary EW(H$\beta$)
between 0.95 and 1.05 times its nominal value. As for the extinction,
we assume that the extinction 
coefficient $C$(H$\beta$)$_{\rm stars}$ for the stellar
light is the same as $C$(H$\beta$)$_{\rm gas}$ for the ionized gas, and we vary
both in the range (0.8-1.2)$\times$$C$(H$\beta$), where $C$(H$\beta$) is the 
extinction coefficient derived from the observed hydrogen Balmer decrement. 
For each LCG, we run 
10$^4$ Monte Carlo models varying $t$(young), $t_1$,
$t$(old), $b$, and $T_e$(H$^+$) randomly in a large range, and EW(H$\beta$) and
$C$(H$\beta$)$_{\rm gas}$ in a relatively smaller range because
the latter quantities are more directly constrained by observations. The best
modeled SED is found from $\chi ^2$ minimization of the deviation between the
modeled and the observed continuum in five wavelength ranges which are free 
of the emission lines.

\begin{figure*}
\figurenum{9}
\hbox{\includegraphics[angle=-90,width=0.49\linewidth]{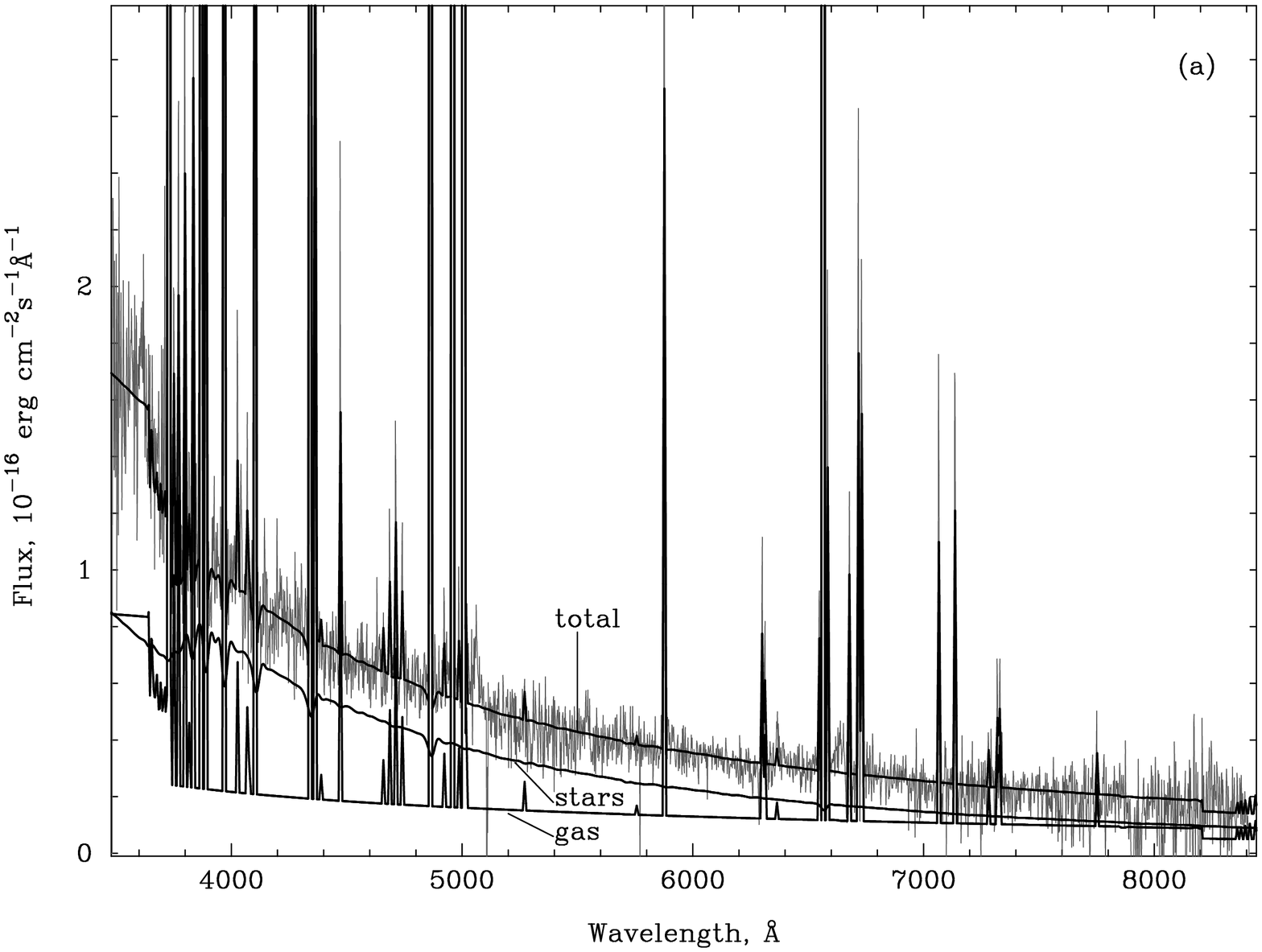} 
\includegraphics[angle=-90,width=0.49\linewidth]{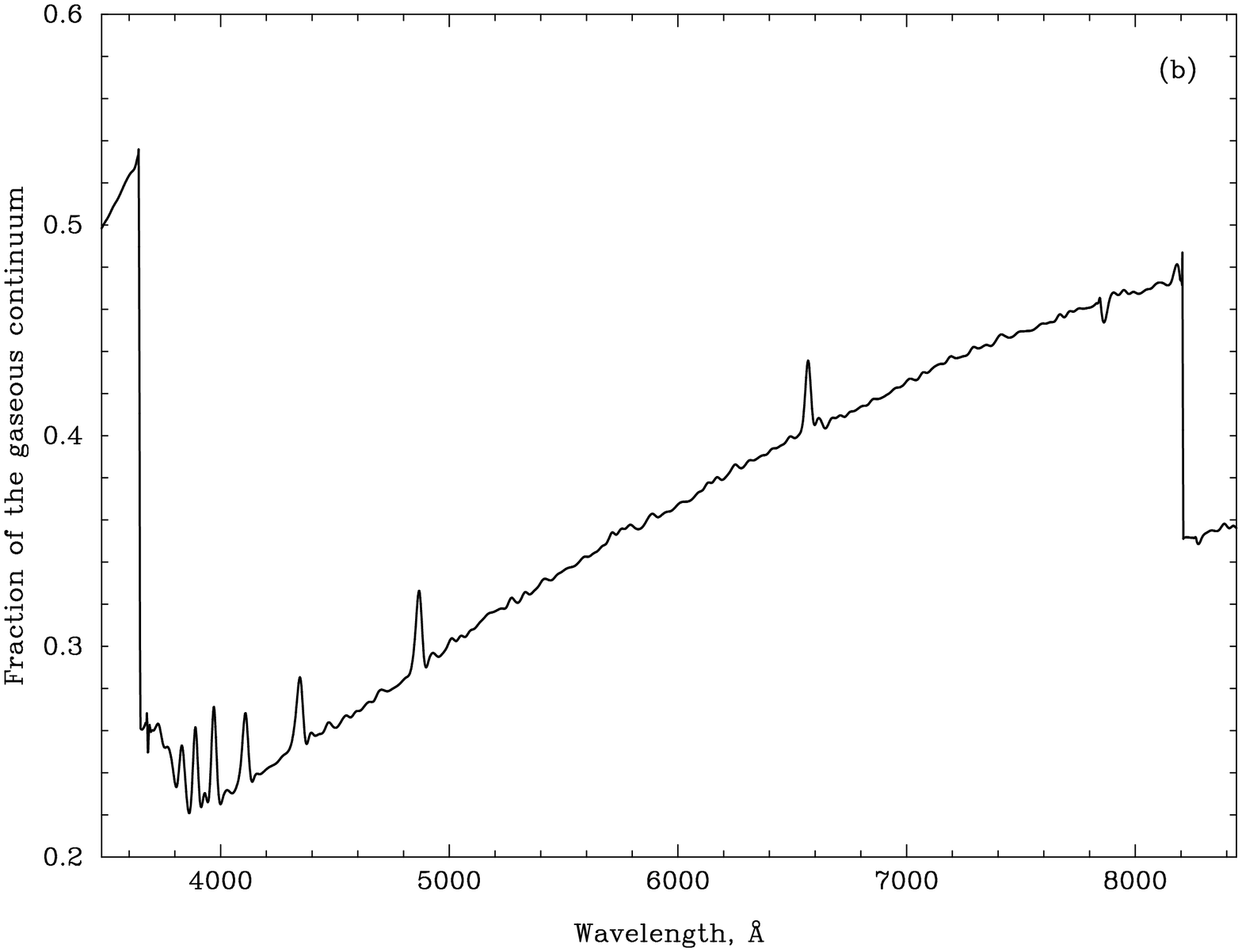}}
\figcaption{(a) Best-fit model SED (solid black line labeled ``total'') 
superposed on the redshift- and extinction-corrected spectrum of the LCG
SDSS J0851+5840 (grey line). The separate contributions from the stellar and 
ionized gas components are shown by black solid
lines and labeled ``stars'' and ``gas'', respectively. (b) Gaseous 
emission fraction vs. wavelength
for the modeled spectrum of LCG SDSS J0851+5840 with EW(H$\beta$) = 303\AA. 
The two jumps seen at $\sim$ $\lambda$3660\AA\ and $\sim$ $\lambda$8200\AA\ 
are due respectively to 
the hydrogen Balmer and Paschen discontinuities in the ionized gas emission.
\label{fig9}}
\end{figure*}

In Fig. \ref{fig9}a we show the SED fitting of the spectrum
of the galaxy J0851+5840 with EW(H$\beta$) = 303\AA. The total modeled spectrum
is labeled ``total'' while the stellar and gaseous spectra are labeled
``stars'' and ``gas'', respectively. The model SED fits very well the observed 
spectrum of the galaxy. It is seen that the contribution of
gaseous emission is high because of the high EW(H$\beta$). The fraction of the
gaseous emission as a function of wavelength in the spectrum of the same
galaxy is shown in Fig. \ref{fig9}b. The two steep jumps seen at 
$\sim$ $\lambda$3660\AA\ and $\sim$ $\lambda$8200\AA\ are respectively due to the Balmer and
Paschen discontinuities of the ionized gas emission. Between 
these jumps,
the fraction of gaseous emission increases monotonously from $\sim$ 20\% to $\sim$ 50\%
making the spectrum flatter than a pure stellar
emission spectrum. Neglecting the contribution of this gaseous emission would have resulted in an  
artificial increase of the galaxy mass. For the stellar mass determination, we use the
SED labeled ``stars'' in Fig. \ref{fig9}a, which is considerably below the SED labeled
``total''. 

\begin{figure*}
\figurenum{10}
\hbox{\includegraphics[angle=-90,width=0.49\linewidth]{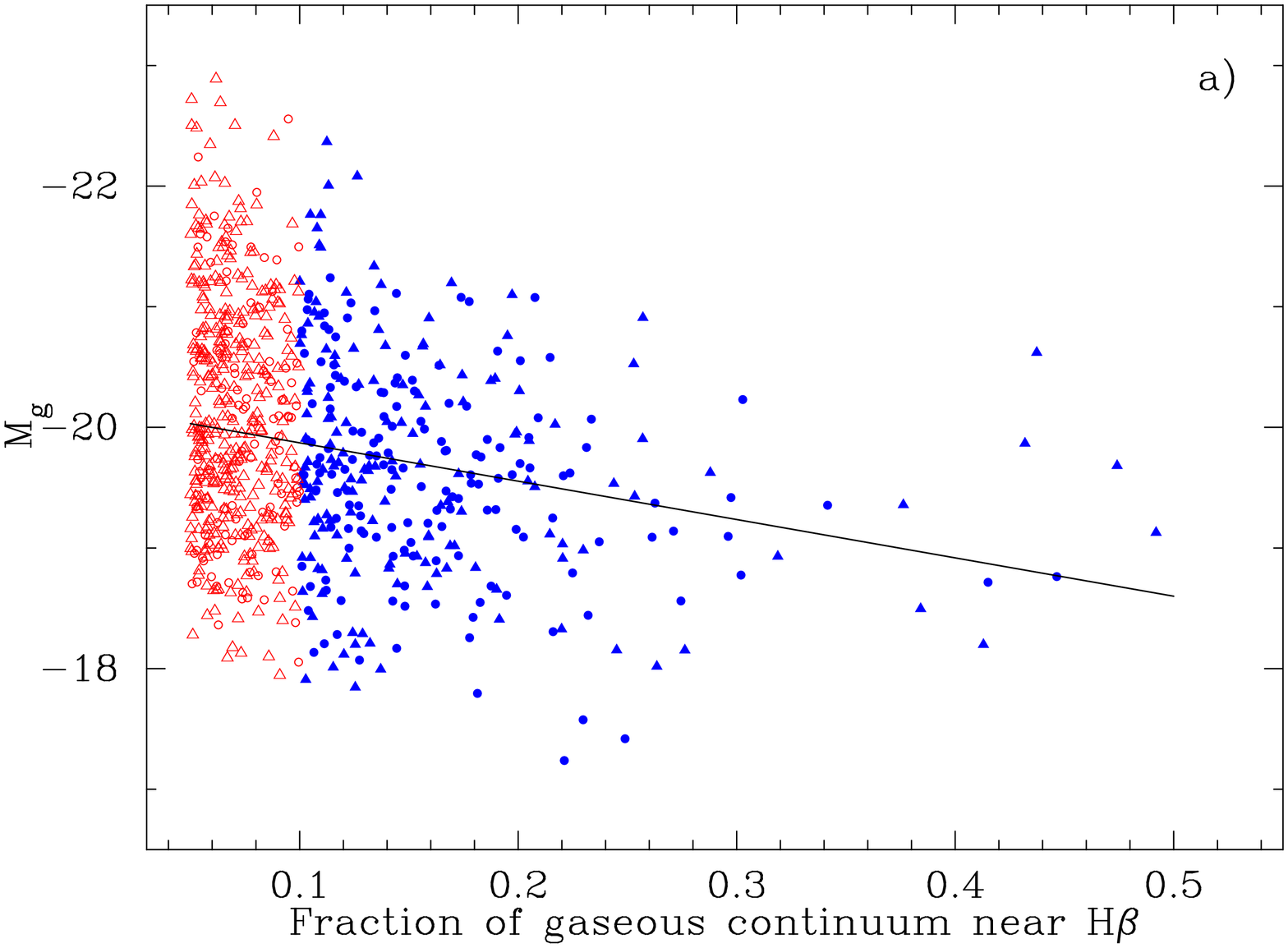} 
\includegraphics[angle=-90,width=0.49\linewidth]{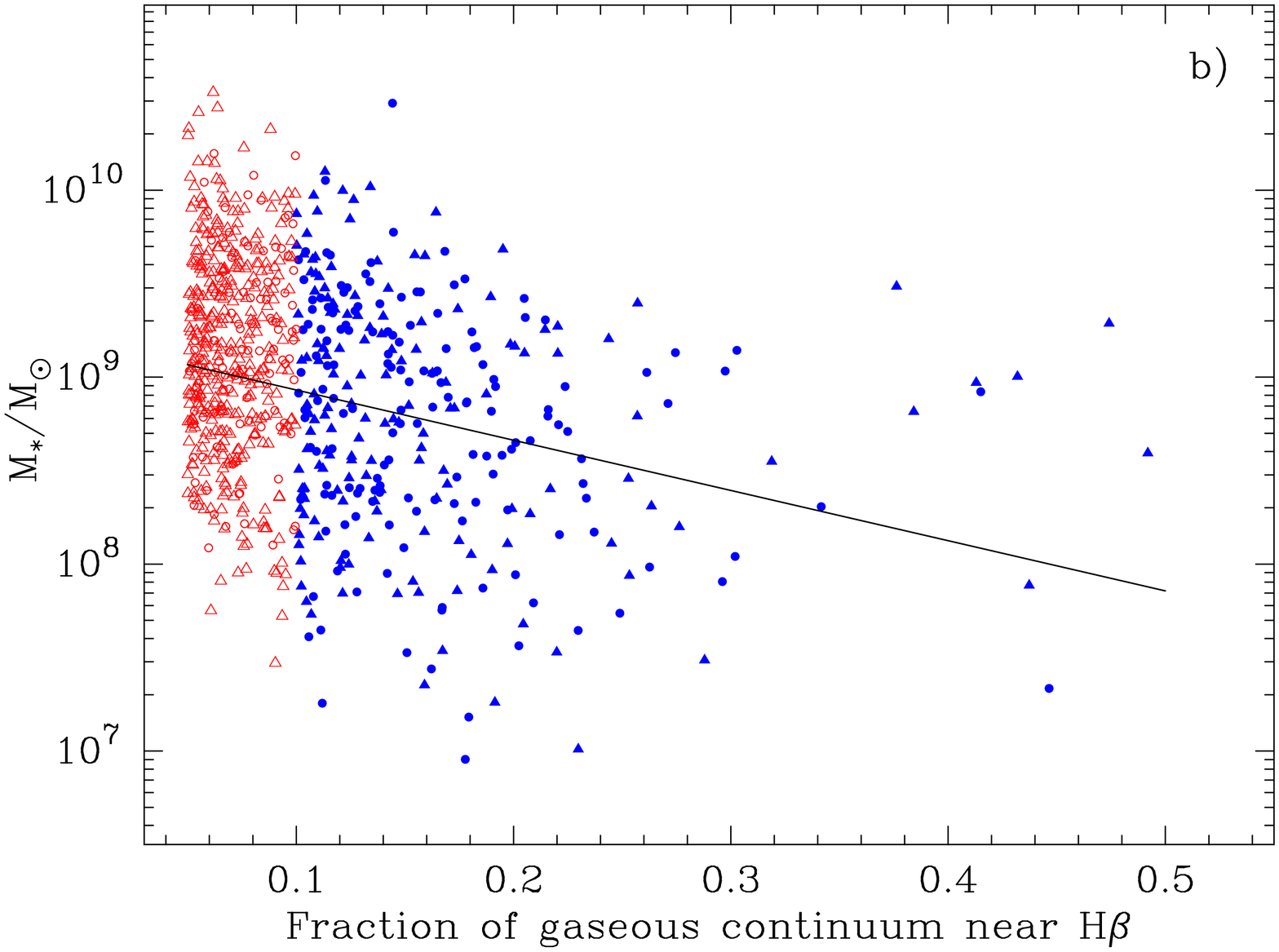}}
\figcaption{(a) The absolute magnitude $M_g$ vs. the fraction of
gaseous continuum near the H$\beta$ emission line. (b) The total stellar mass
vs. the fraction of gaseous continuum near the H$\beta$ emission line.
In both panels, the straight line represents the linear 
best likelihood fit. \label{fig10}}
\end{figure*}

In Fig. \ref{fig10}, we show the dependence of (a) the absolute $g$ magnitude
$M_g$ and (b) of the total stellar mass $M_*$ on the fraction of
gaseous continuum near the H$\beta$ emission line. The solid lines are 
the best linear likelihood fits to the data. It is seen that galaxies with
high EW(H$\beta$) (filled symbols) are on average fainter and less massive 
than galaxies with low EW(H$\beta$) (open symbols).

\subsection{Masses of young and old stellar populations}

From the SED fits of the spectra of all the 803 LCGs in our sample, we find the median
total stellar mass to be $M_*$ = 1$\times$10$^{9}$ $M_\odot$ 
(Table \ref{tab1}). This value is a factor of $\sim$ 3 lower than the 
median mass 3$\times$10$^{9}$ $M_\odot$ found by 
\citet{C09} for their sample of green pea galaxies.
As noted before, these star-forming green pea galaxies
with their round morphology and high EW(H$\beta$) are similar to
the LCGs in our subsample 1 which is characterized by an even smaller
median mass of 7$\times$10$^{8}$ $M_\odot$ (Table \ref{tab1}).
We believe the factor of $\sim$4 in median mass between the \citet{C09} green pea sample and our 
subsample 1 is primarily due to the fact that \citet{C09} did not subtract 
the gaseous continuum emission from the total galaxy emission in deriving their stellar masses. A more detailed comparison between 
stellar masses derived by us and by \citet{C09} is given in Section 5.3. 

As for the young stellar population in LCGs, its median mass for 
the whole sample is $M$(young) = 3$\times$10$^{7}$ $M_\odot$, i.e. 
a factor of $\sim$ 30 below the median mass of the old stellar population.
 There is a
small increase of the median mass of the young stellar population by a factor
of $\sim$ 1.4 from subsamples 1 and 2 to subsamples 3 and 4 (Table \ref{tab1}).
For comparison, the median total stellar
mass of LCGs in subsample 1 is $\sim$ 2 times smaller than that in 
subsample 4.

\begin{figure*}
\figurenum{11}
\hbox{\includegraphics[angle=-90,width=0.49\linewidth]{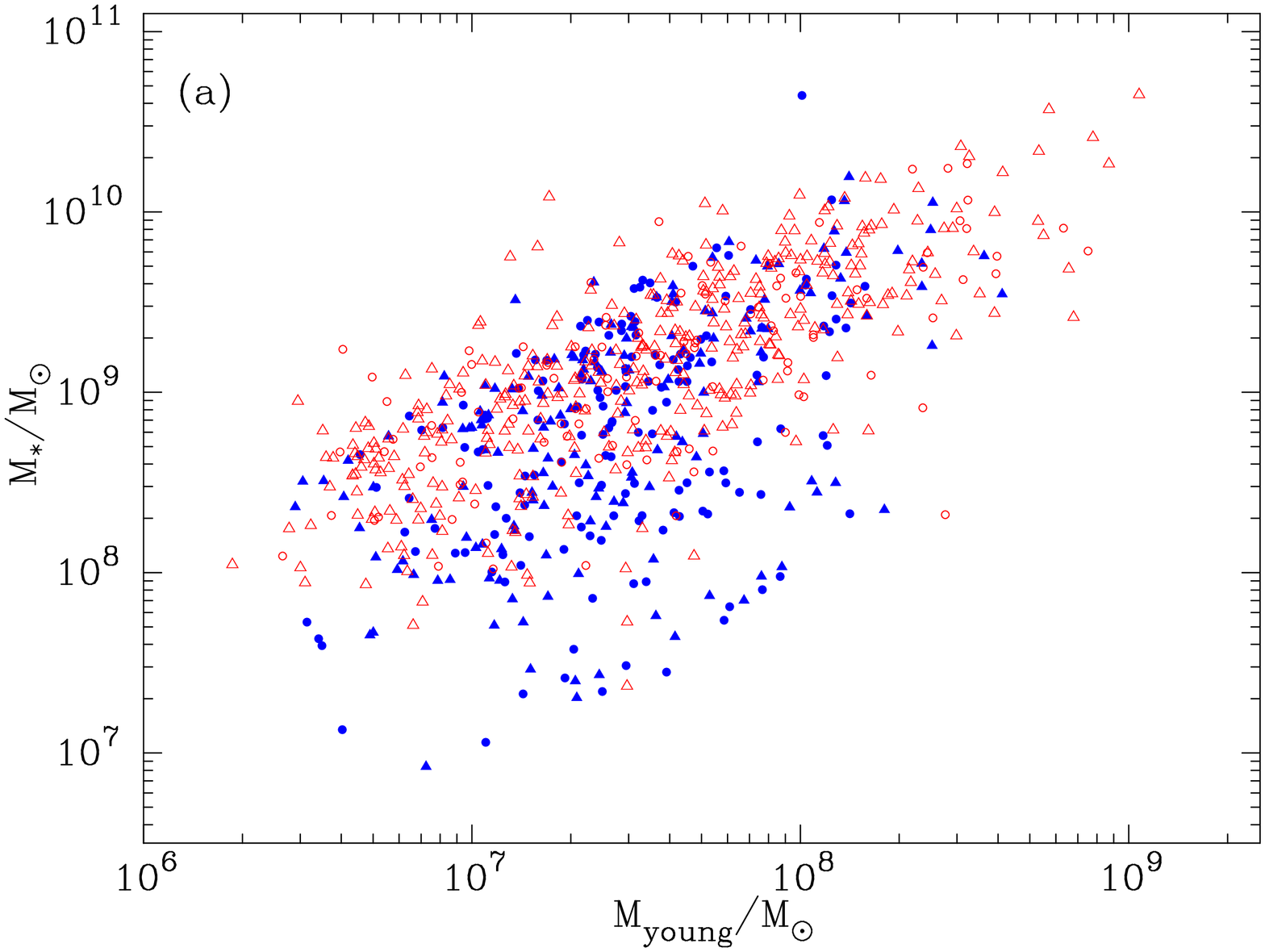} 
\includegraphics[angle=-90,width=0.49\linewidth]{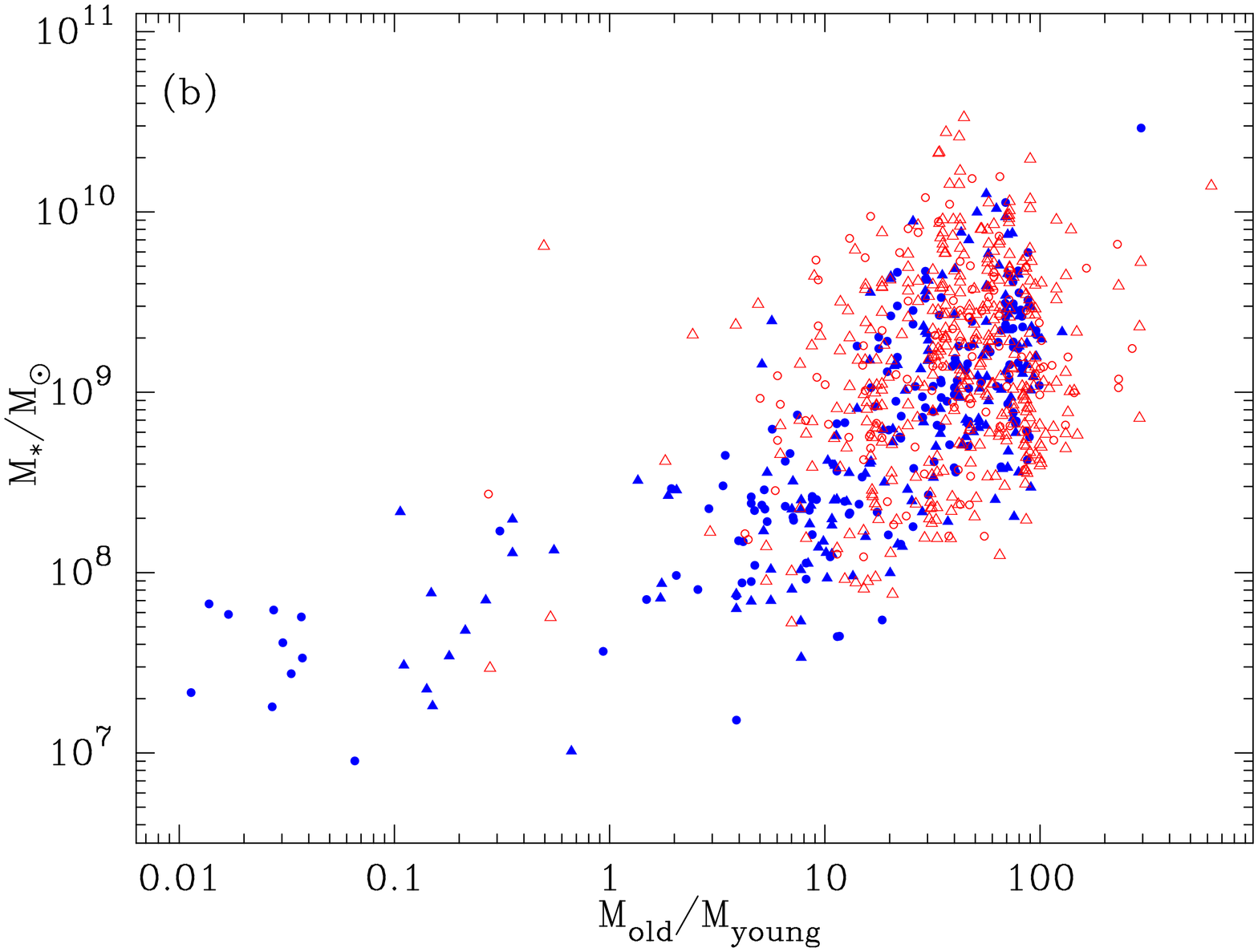}}
\figcaption{(a) The total stellar mass $M_*$ vs. the 
young stellar population mass $M$(young). (b) The total stellar mass
vs. the young-to-old stellar population mass ratio. The masses of 
massive galaxies are dominated by an old stellar population, while the masses 
of low-mass galaxies are dominated by a young stellar population. \label{fig11}}
\end{figure*}

Fig. \ref{fig11}a shows the dependence of the total stellar mass 
$M_*$ of LCGs on the mass $M$(young) of the young stellar population.
We note that the total stellar mass is in the range 
10$^7$ -- 5$\times$10$^{10}$ 
$M_\odot$, while the mass of the young stellar population is in the range
2$\times$10$^6$ -- 10$^9$ $M_\odot$. 
Evidently, there is an increase of the mass of the 
young stellar population with increasing
total stellar mass of the galaxy. This suggests that more massive
galaxies contain more regions of ongoing star formation. 
Fig. \ref{fig11}b shows the dependence of the total stellar mass on the
ratio $b$ of the mass of the old stellar population to that of the young 
stellar population. It is seen that 
$M$(old)/$M$(young) increases with increasing $M_*$, reaching
the high values of several hundreds, predominantly 
in galaxies with low EW(H$\beta$) (open symbols). 
On the other hand, there is an appreciable number of LCGs with
high EW(H$\beta$) and low $M$(old)/$M$(young) (filled symbols), implying that 
less massive dwarf galaxies (10$^7$$M_\odot$ $\la$ $M_*$ $\la$ 
10$^8$$M_\odot$) are also less evolved, both in terms of their stellar 
populations and metallicities.

In the fitting analysis and in the determination of global characteristics
of LCGs such as luminosities and masses, we have 
made no correction for aperture effects.
Yet, these effects can be important, especially for galaxies at
low redshifts, which may have angular 
sizes larger than the 3\arcsec\, diameter 
of the fibers with which the SDSS spectra were obtained. 
We estimate these aperture corrections below. 
Assuming that the spectral energy distribution in the galaxy inside
and outside the round aperture are the same, the aperture correction 
factor can be written as
\begin{equation} 
f(apcorr) = 10^{0.4(r(3\arcsec) - r({\rm tot}))}, \label{eqapcor}
\end{equation}
where $r$(3\arcsec) and 
$r$(tot) are respectively the SDSS $r$ apparent magnitude inside the
3\arcsec\ round aperture and the modeled integrated SDSS $r$ 
apparent magnitude of the galaxy as
derived from the SDSS images. Both $r$(3\arcsec) and  $r$(tot)
are taken from the SDSS DR7 database. Using Eq. \ref{eqapcor},  
we found the median flux correction factor  $f(apcorr)$ to vary from 
$\sim$ 1.4 for subsample 1 to $\sim$ 1.8 for subsample 4. However,
for comparison purposes, we have decided 
not to apply these aperture corrections to the LCG sample,   
as they are not taken into account in other studies. Furthermore, the
assumption that the spectral energy distributions of the galaxy inside
and outside the slit are the same may not be valid. Thus, we should 
keep in mind that
the true global characteristics for LCGs, 
such as luminosity, mass or star formation rate,
may be larger than those derived here from the galaxy spectra, especially
for nearby galaxies.

\subsection{Comparison of stellar masses and oxygen abundances 
of green pea galaxies derived by us and by \citet{C09}}

\begin{figure*}
\figurenum{12}
\hbox{\includegraphics[angle=-90,width=0.49\linewidth]{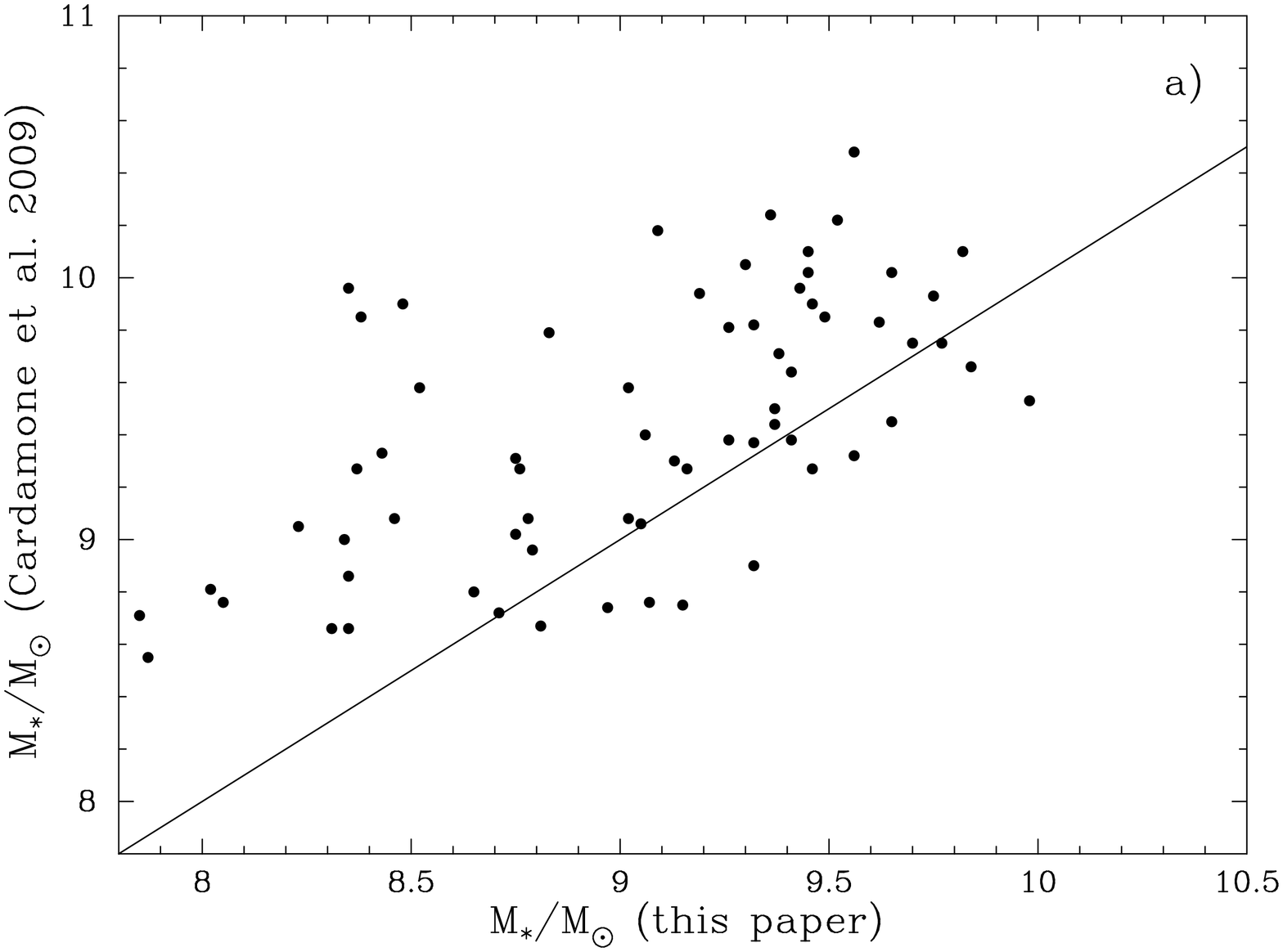} 
\includegraphics[angle=-90,width=0.49\linewidth]{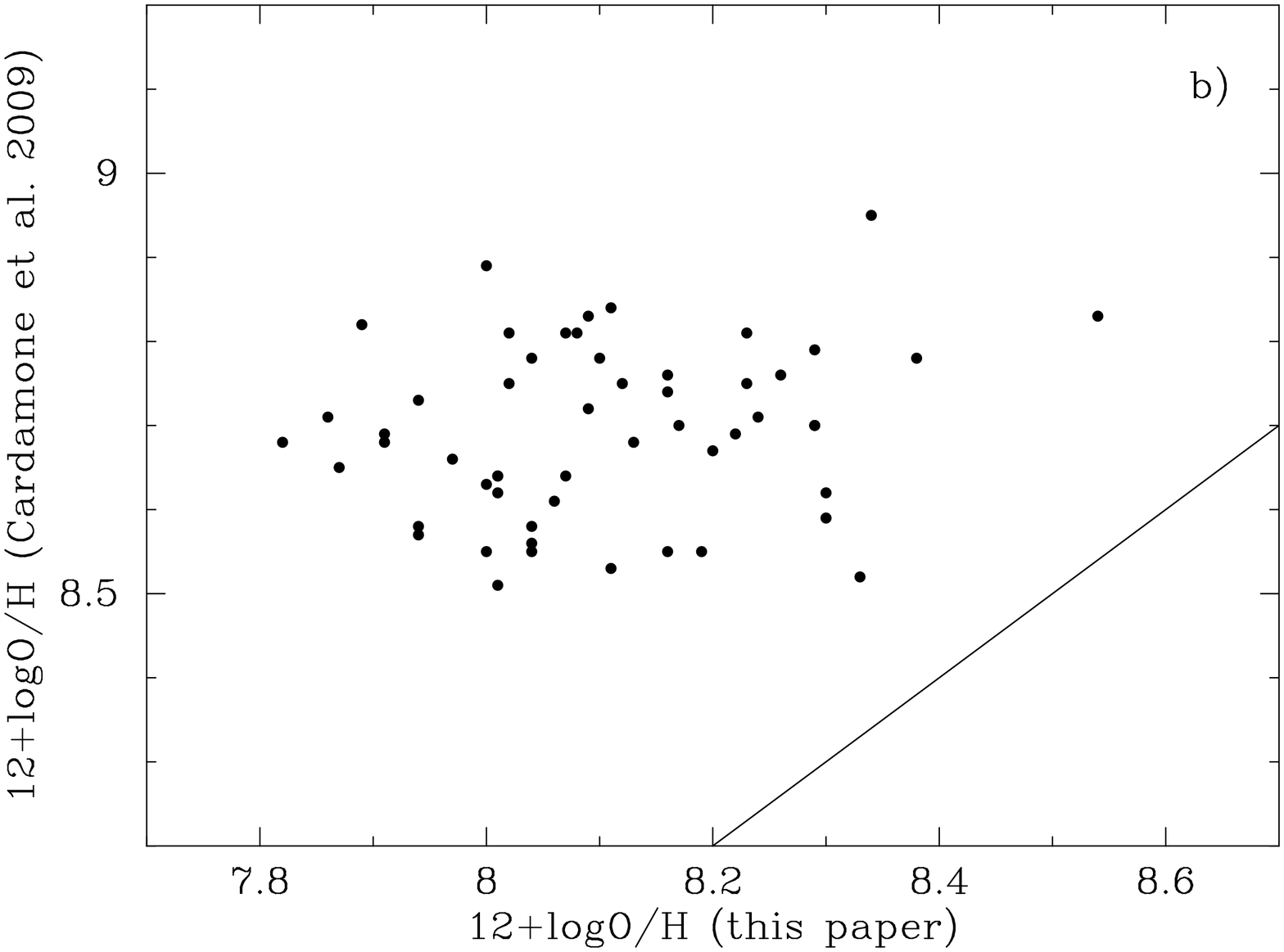}}
\figcaption{Comparison of: (a) stellar masses and (b) oxygen
abundances derived for green pea galaxies by \citet{C09} and
in this paper. Solid lines in both panels are lines of equal values.
\label{fig12}}
\end{figure*}

As discussed before, there are 66 green pea galaxies from \citet{C09} that are
included in the LCG sample.
We compare here their stellar masses and oxygen abundances derived by 
us to those obtained by \citet{C09}. 
We show in Table \ref{tab2} the SDSS ID number of each galaxy,
its equatorial coordinates, 
its stellar masses and oxygen abundances derived by
\citet{C09} and in this paper. An examination of the Table shows that the
stellar masses of green pea galaxies derived in this paper are 
systematically lower than those obtained by \citet{C09}. This 
effect is also clearly seen in Fig. \ref{fig12}a. 
As already discussed, this systematic offset is due to the fact that,
in fitting the SED,  
we subtract the contribution of gaseous continuum emission, which
results in lower stellar masses. This effect is
largest for lower-mass galaxies with log $M_*$/$M_\odot$ $\la$ 8.5. 
These are mostly 
galaxies with high EW(H$\beta$) in subsample 1.

Fig. \ref{fig12}b shows the comparison of oxygen abundances 
for the same galaxies. 
Again, the \citet{C09} abundances are 
systematically larger ($\sim$ 0.6 dex in 12+logO/H) than ours. 
This systematic offset 
is a consequence of the different abundance determination methods used,
the more uncertain strong-line method by \citet{C09}, and the 
more reliable direct method by us.  
 We note that nearly all abundance 
determinations in high-redshift galaxies and in the majority of nearby SDSS 
star-forming galaxies are based on strong-line methods. 
The more uncertain oxygen abundances that result from these methods 
should not be used in galaxy evolution studies 
without a careful analysis of their errors. 

\section{Younger starbursts are hotter and denser than older ones}

\begin{figure*}
\figurenum{13}
\hbox{\includegraphics[angle=-90,width=0.49\linewidth]{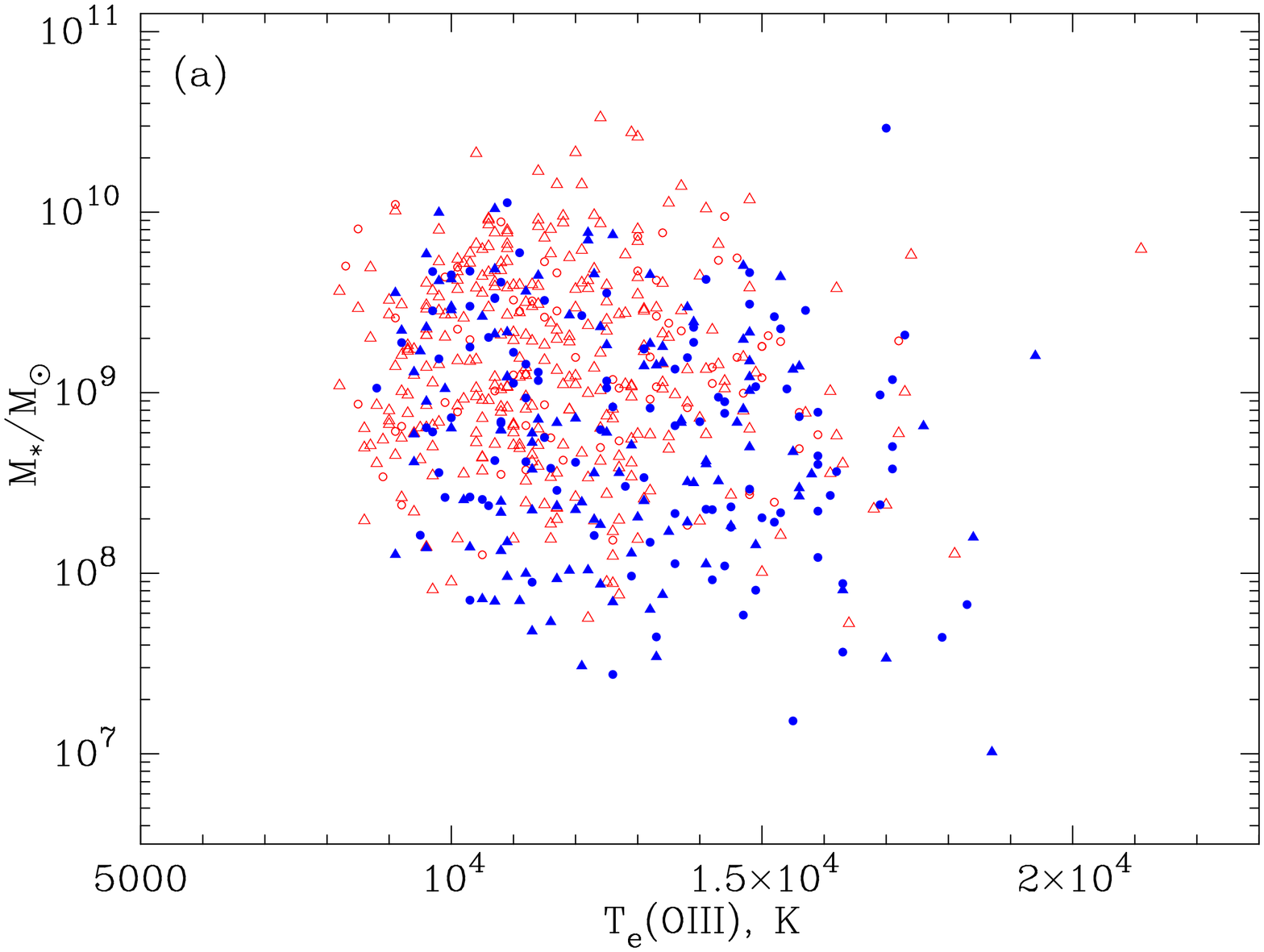} 
\includegraphics[angle=-90,width=0.49\linewidth]{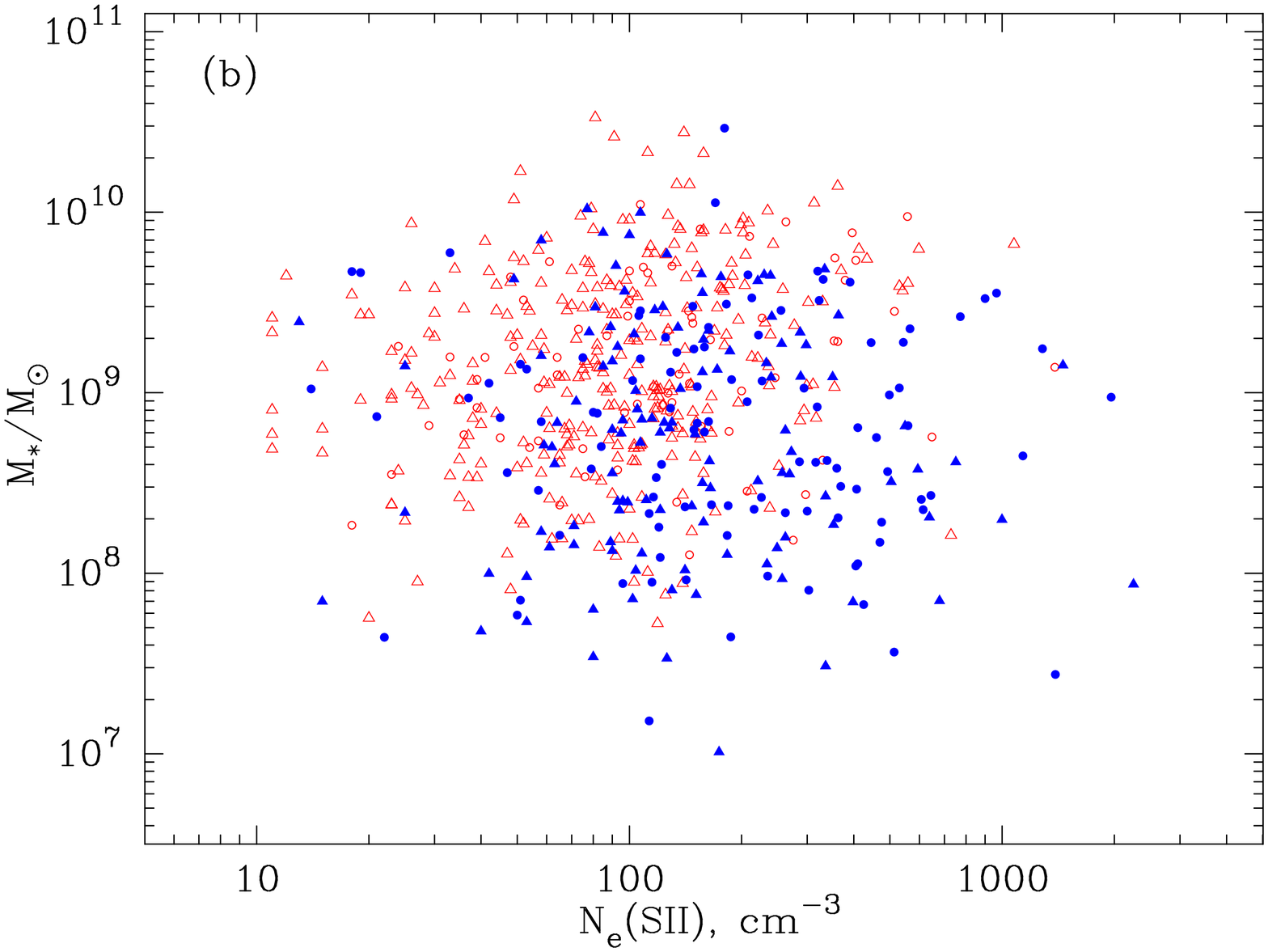}}
\figcaption{The total stellar mass $M_*$ vs. (a) 
the electron temperature $T_e$(O {\sc iii}) and (b) the electron
number density $N_e$(S {\sc ii}) of the ionized gas. Both
$T_e$(O {\sc iii}) and $N_e$(S {\sc ii}) are systematically higher in 
low-mass galaxies with high EW(H$\beta$). \label{fig13}}
\end{figure*}

We investigate here how the properties of the luminous H {\sc ii}
regions in LCGs depend on the stellar mass of the galaxy. In particular, 
we examine the dependence 
of the electron temperature $T_e$(O {\sc iii}) and the
electron number density $N_e$(S {\sc ii}) on the total stellar mass $M_*$
of the galaxy (respectively Figs. \ref{fig13}a and \ref{fig13}b).
We find that both the electron temperature and the electron number density
are systematically higher in galaxies with higher EW(H$\beta$). These galaxies
also have systematically lower masses. The median electron temperatures 
and electron number densities of the galaxies in the different LCG subsamples 
vary from 13~100K and 180 cm$^{-3}$ in subsample 1, 
to 11~700K and 90 cm$^{-3}$ in subsample 4. 

Young stellar populations in galaxies with higher EW(H$\beta$) 
have younger ages $t$(young) (Table \ref{tab1}). Thus, we conclude that the 
younger starbursts in our LCGs are systematically hotter and denser. 

\section{Mass-metallicity relation}

\begin{figure*}
\figurenum{14}
\hbox{\includegraphics[angle=-90,width=0.49\linewidth]{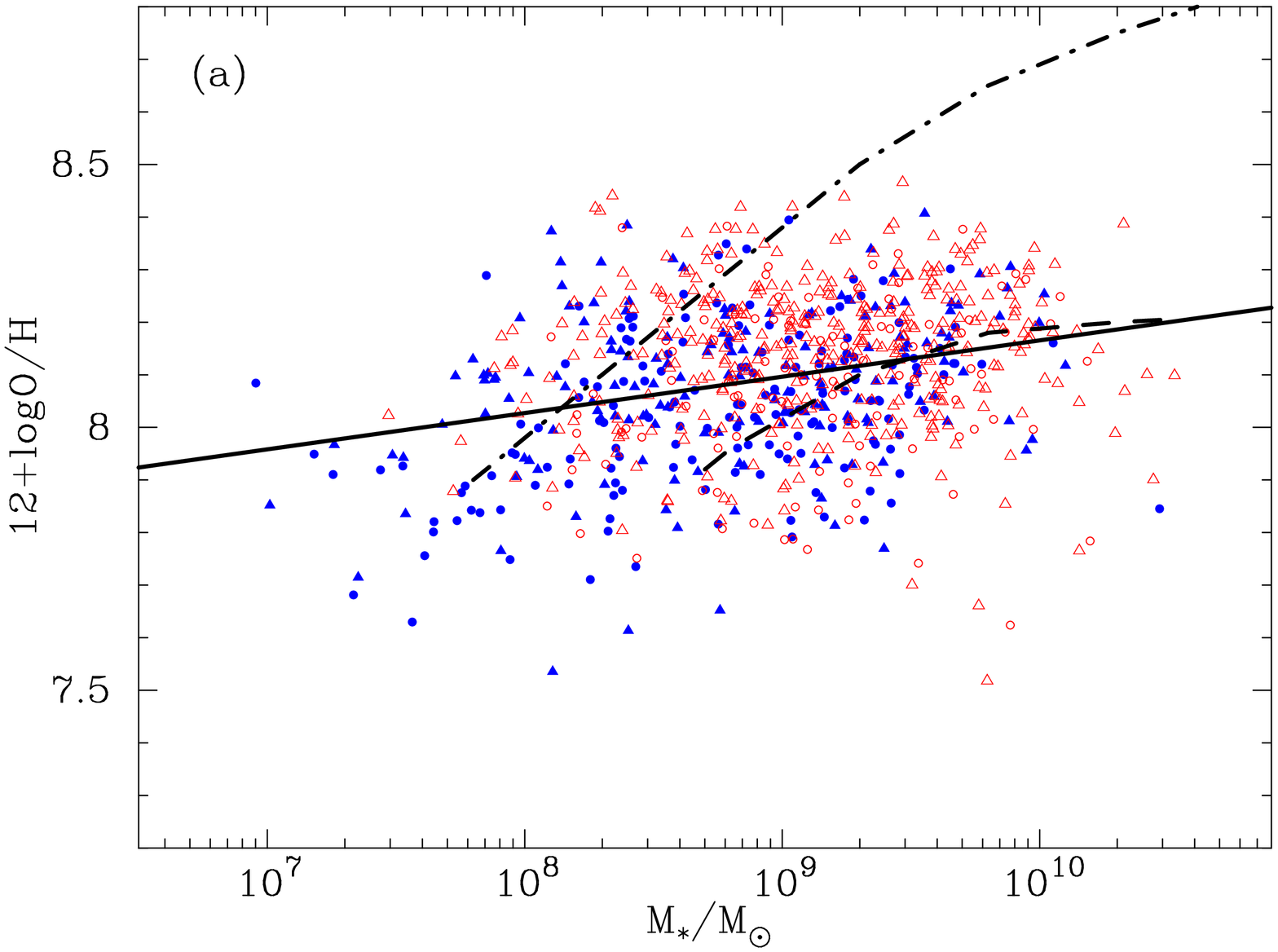} 
\includegraphics[angle=-90,width=0.49\linewidth]{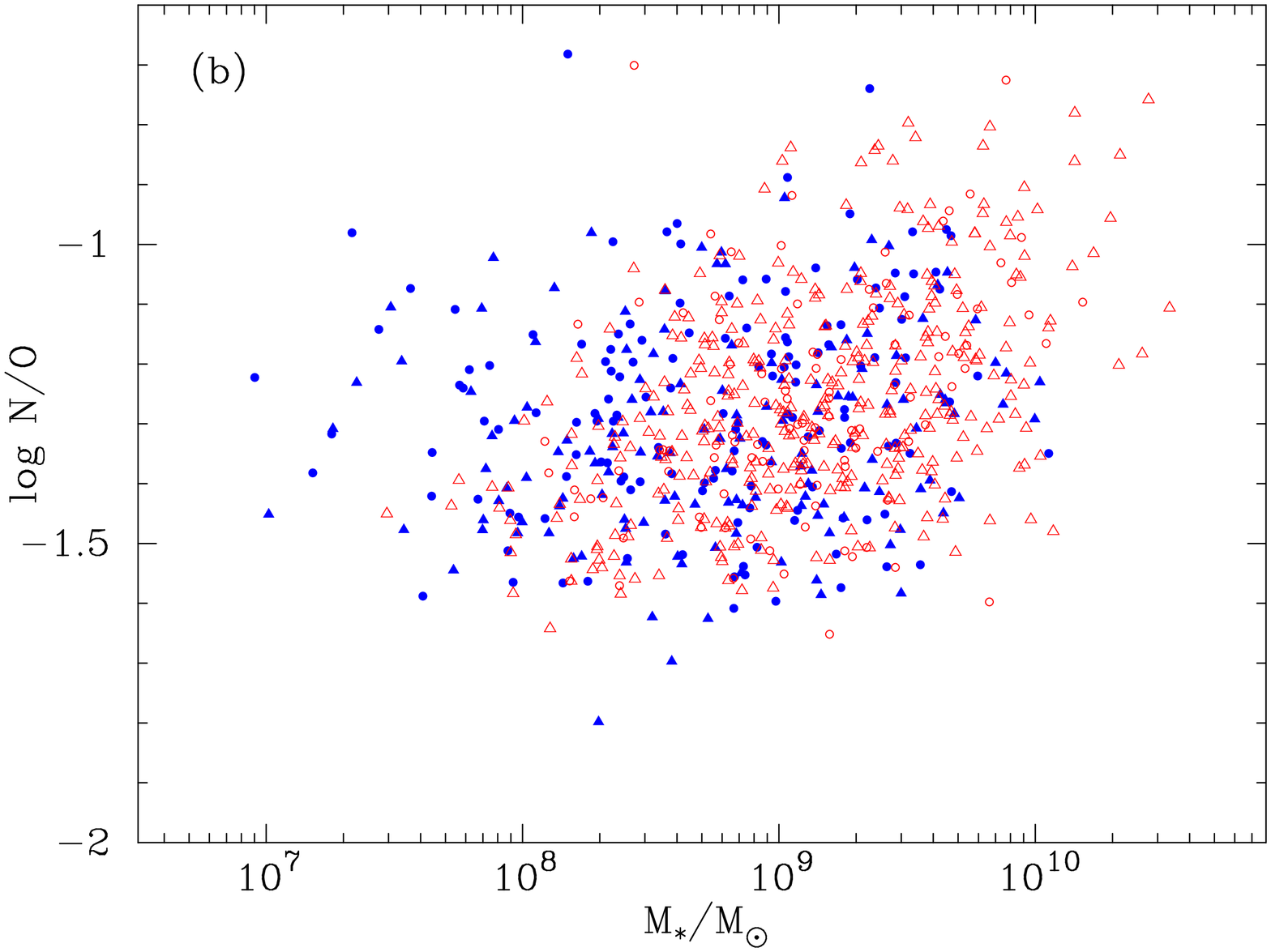}}
\figcaption{(a) Oxygen abundance 12 + log O/H vs. stellar mass 
$M_*$ for the LCG sample. 
The symbols are the same as in as in Fig. \ref{fig2}.
The linear least-squares best fit to the data is shown
by the solid line. 
The fits to the green pea sample and star-forming SDSS galaxies of 
\citet{A10} are shown respectively by dashed and dash-dotted lines.
(b) The N/O vs. $M_*$ relation for the LCG sample.
The symbols are the same as in Fig. \ref{fig2}. \label{fig14}}
\end{figure*}

In Fig. \ref{fig14}a,b  we show the mass-metallicity relation for LCGs. 
It is seen that galaxies with 
lower mass have systematically lower oxygen abundances. 
These lower-mass galaxies are
primarily from subsamples 1 and 2 with high EW(H$\beta$), while LCGs with lower
EW(H$\beta$) are higher-mass objects with higher oxygen abundance.
Although there is a large scatter , the
data can be fitted by a linear relation, shown in Fig. \ref{fig14}a
by a solid line
\begin{equation}
12+\log{\rm O/H} = (7.475\pm 0.426) + (0.069\pm 0.047)\times \log M_*, \label{eg2}
\end{equation}
where $M_*$ is in solar units. 
We also show by a dashed line in Figure \ref{fig14}a 
the polynomial fit obtained by \citet{A10} to the green pea galaxy 
sample of \citet{C09}. The
difference between our fit and that of \citet{A10} is due to the fact that 
we take into account the gaseous continuum emission and thus derived
lower stellar masses producing an offset by a factor of $\sim$ 4 
in the galaxy mass between our data and \citet{A10} fit.
Furthermore, because of 
a small stellar mass range and a large dispersion in oxygen abundances, 
there is no strong evidence from Fig. 3 of \citet{A10}
that the data for green pea galaxies is better fit  
by a polynomial than a line.
A linear fit to the \citet{A10} data would result in a better agreement with
the fit to LCGs derived in this paper. 

We show by a dash-dotted line the fit of \citet{A10}
to a sample of star-forming galaxies assembled from the SDSS. This relation 
is considerably steeper than the one obtained for LCGs. 
The difference is probably not due to the determination of the stellar
masses of the SDSS star-forming galaxies.
The equivalent width of their H$\beta$ emission line is generally small, 
so that ionized gas emission is small. 
However, the oxygen 
abundances of these galaxies are derived using strong-line empirical methods 
which produce a systematic offset as compared to 
the oxygen abundances derived by 
the direct method. That is probably why  
the relations for SDSS 
star-forming galaxies and LCGs differ.

The relation shown by the solid line in Fig. \ref{fig14}a  
is also at odds with the finding of 
\citet{C09} for their green pea sample. Those authors 
did not find a clear evidence for the dependence
of the oxygen abundance on stellar mass. This is likely because of the way 
they derive the metallicities and masses of their galaxies. 
As discussed before, oxygen abundances were derived 
by using the more uncertain strong-line technique, 
and stellar masses were determined without subtracting  
the gaseous continuum emission from the observed SED. In that way, both 
oxygen abundances and galaxy stellar masses are
overestimated. 
Another reason for why a mass-metallicity relation 
was not seen by \citet{C09} is probably because the 10$^8$ - 10$^{10.5}$ 
$M_\odot$ mass range of their sample is smaller than the 
10$^7$ - 10$^{10.5}$ $M_\odot$ mass range of our LCG sample.

In contrast to Fig. \ref{fig14}a, Fig. \ref{fig14}b 
shows that the N/O abundance ratio for LCGs
with EW(H$\beta$) $\geq$ 100\AA\ (filled symbols)
is independent on stellar mass, implying a primary origin for 
nitrogen in these galaxies. As for LCGs with EW(H$\beta$) $<$ 100\AA\ 
(open symbols), 
there is a dependence of N/O ratio on the total galaxy mass,
implying some secondary nitrogen production in the highest-mass galaxies.

\section{Star formation rate}

One of the most 
important characteristics governing a galaxy's evolution is its star
formation rate. We derive the star formation rate SFR(H$\alpha$), using the 
extinction-corrected luminosity $L$(H$\alpha$) of the 
H$\alpha$ emission line and the star formation rate 
SFR(FUV), using the extinction-corrected GALEX FUV luminosity 
$L$(FUV), and relations given by \citet{K98}
\begin{equation}
{\rm SFR}({\rm H}\alpha)=7.9\times10^{-42} L({\rm H}\alpha), \label{eq3}
\end{equation}
\begin{equation}
{\rm SFR(FUV)}=1.4\times10^{-28} L({\rm FUV}). \label{eq4}
\end{equation}
In equations \ref{eq3} and \ref{eq4}, SFR is in units of 
$M_\odot$ yr$^{-1}$, $L$(H$\alpha$) in erg s$^{-1}$
and $L$(FUV) in erg s$^{-1}$ Hz$^{-1}$. To correct $L$(FUV) for extinction
we adopt the reddening law of \citet{C89} with $R$ = $A(V)$/$E(B-V)$  = 3.1 and
the extinction coefficient $C$(H$\beta$) derived from the optical SDSS
spectrum. Then the extinction correction factor for $L$(FUV) is equal to
2.512$^{A({\rm FUV})}$ with $A$(FUV) = 8.24$\times$$E(B-V)$ 
= 5.58$\times$$C$(H$\beta$).

The GALEX FUV fluxes are available for 675 galaxies out of the 803 LCGs in our 
sample. In Table \ref{tab1},
we give the median SFR(H$\alpha$) and SFR(FUV) for the 
different subsamples and for
the total sample. It can be seen that 
the SFR(H$\alpha$) and SFR(FUV) are in good agreement 
and that their median values do not change appreciably from one subsample to 
the next, being    
$\sim$ 4 $M_\odot$ yr$^{-1}$. Because of the 
good agreement, we will consider
below only SFRs derived from the H$\alpha$ luminosity.

\begin{figure*}
\figurenum{15}
\hbox{\includegraphics[angle=-90,width=0.49\linewidth]{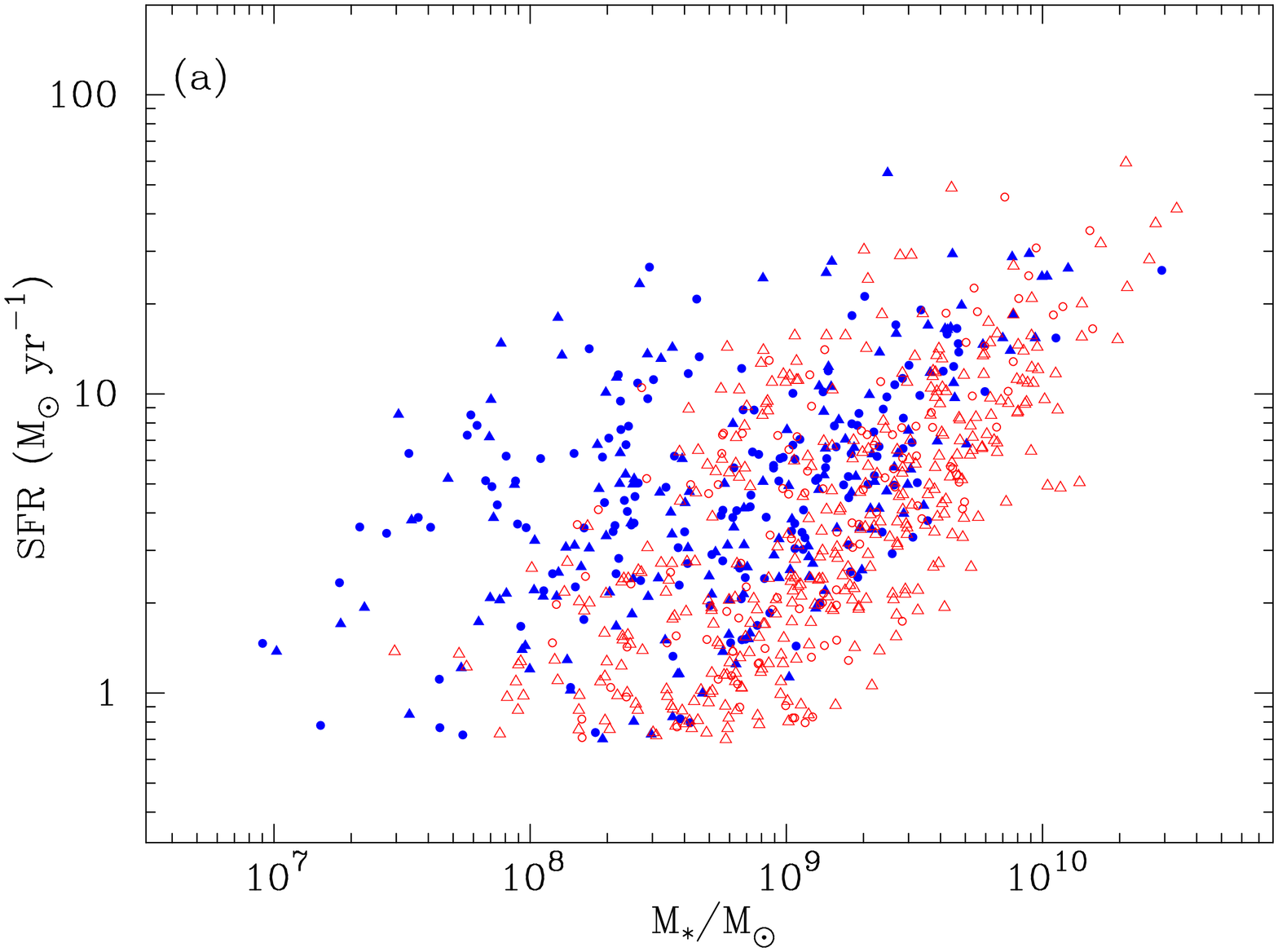} 
\includegraphics[angle=-90,width=0.49\linewidth]{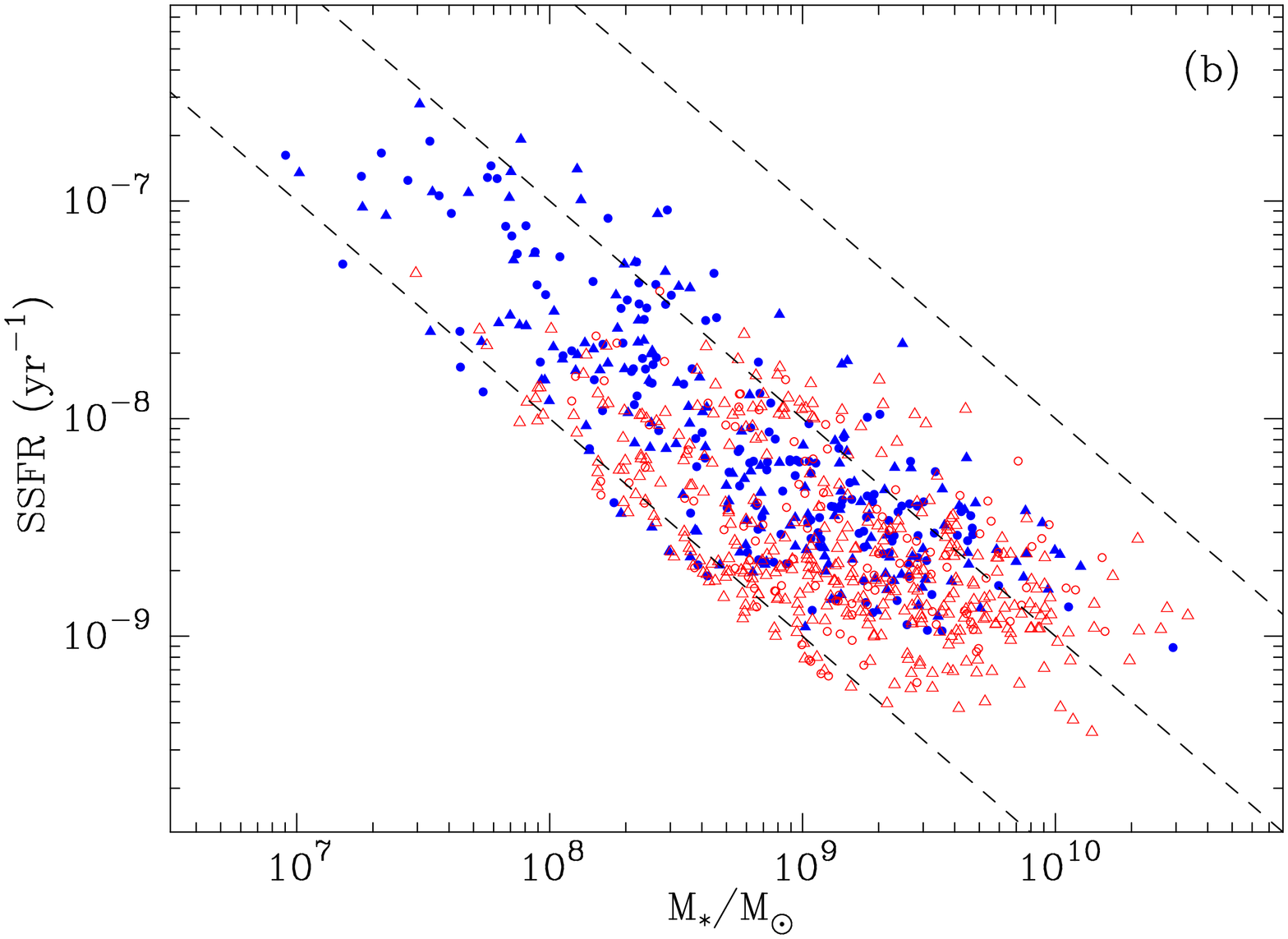}}
\figcaption{(a) The star formation rate SFR(H$\alpha$)  
and (b) the specific star formation rate SSFR(H$\alpha$) vs. the total 
stellar mass $M_*$. The dashed lines in (b) correspond, 
from left to right,  
to loci with SFR = 1, 10 and 100 $M_\odot$ yr$^{-1}$, respectively. 
\label{fig15}}
\end{figure*}

In Fig. \ref{fig15}a we show the dependence of SFR(H$\alpha$) 
on the total stellar mass $M_*$. The star formation rate for LCGs is in 
the range 0.7 - 60 $M_\odot$.
The median values in different subsamples are in the narrow range 
3.4 -- 5.0 $M_\odot$ yr$^{-1}$ (Table \ref{tab1}). 
Thus, there is no significant  
difference in the average SFR for galaxies with different EW(H$\beta$). 
However, we note the increase of SFR(H$\alpha$) with increasing 
total stellar mass in galaxies from
subsamples 3 and 4 (galaxies with EW(H$\beta$) $<$ 100\AA\ shown in 
Fig. \ref{fig15}a by open symbols), while there is a less strong 
dependence for galaxies from subsamples 1 and 2 (galaxies with 
EW(H$\beta$) $\ge$ 100\AA\ shown in Fig. \ref{fig15}a by filled symbols).
The range of SFRs in LCGs is comparable to that in the intermediate-redshift
star-forming galaxies \citep{H05,K07} and in LBGs \citep{P01}, but is $\sim$
10-100 times higher than in typical nearby BCDs.
For example, the SFR in the prototype BCD I Zw 18 is 
0.1 $M_\odot$ yr$^{-1}$ \citep{T08}.

The dependence of the specific star formation rate SSFR(H$\alpha$) = 
SFR(H$\alpha$)/$M_*$
is shown in Fig. \ref{fig15}b. The SSFR is of interest 
because it is related to the time needed to double the stellar mass of a 
galaxy. We show by dashed lines from left to right the loci 
of constant star formation corresponding to 
1, 10 and 100
$M_\odot$ yr$^{-1}$, respectively. It is seen that there is an 
inverse dependence of the SSFR on the total stellar mass of LCGs. The lowest
SSFRs ($\sim$ 10$^{-9}$ yr$^{-1}$) occur in the
highest-mass galaxies with EW(H$\beta$) $<$ 100\AA, while the highest SSFRs
(up to $\sim$ 10$^{-7}$ yr$^{-1}$) are found almost exclusively in
galaxies with EW(H$\beta$) $\ge$ 100\AA. These values are very high, 
higher than those found by \citet{C09} for their green pea galaxies
(their maximum SSFRs are $\sim$ 10$^{-8}$ yr$^{-1}$), 
mainly because those authors overestimated the galaxy stellar masses.  
They are also considerably higher 
than those found in merging galaxies which lie 
in the 10$^{-10}$ yr$^{-1}$ -- 10$^{-9}$ yr$^{-1}$ range. 
Our SSFR values are comparable to
those found in high-redshift galaxies in the ranges 
$z$ = 4--6 \citep{S09} and 6--8 \citep{SB10}. Thus, our LCGs resemble 
galaxies with
the highest SSFR known.

The quantity 1/SSFR(H$\alpha$) is also a measure of the 
time needed to double the metallicity of a galaxy.
Thus, if a constant star formation rate is assumed, then
the ISM in our galaxies is enriched by heavy elements
on a time scale of $\sim$ 10$^{7}$ - 10$^{9}$ yr, considerably 
smaller than the
Hubble time. It is likely then that star formation in LCGs
proceeds in short bursts of duration 10$^{6}$ - 10$^{7}$ yr. 
Then, the SSFR averaged over time would be much 
lower. Bursts of star formation separated by long quiescent periods of 
several Gyr \citep[not unlike the situation in BCDs, see ][]{T08} are 
also required in the LCGs with the highest SFRs, 
so as not to exhaust the gas supply for 
star formation. Assuming continuous star formation, then with a hydrogen gas
supply of  10$^{10}$ $M_\odot$ and a SFR of 50 $M_\odot$ yr$^{-1}$,
the time for gas depletion is only $\sim$ 2$\times$10$^{8}$ yr. 
Clearly a burst mode is required for the LCGs with the highest SFRs.

\section{Age of stellar populations and the burst mode of star formation 
in LCGs}

While the young stellar population in our galaxies is modeled by a short burst
occurring at $t$(young) $\sim$ 3 - 4 Myr (Table \ref{tab1}), the older stellar 
population is fitted by a model with continuous star formation starting at 
$t_2$ $\equiv$ $t$(old) and ending at $t_1$, with $t_2$ $>$ $t_1$. 
The parameter
$t$(old) sheds light on the evolutionary status of the
galaxy (whether the galaxy is young or old), and helps 
to understand the character of its 
star formation (whether it proceeds in a bursting or continuous manner).

\begin{figure*}
\figurenum{16}
\hbox{\includegraphics[angle=-90,width=0.9\linewidth]{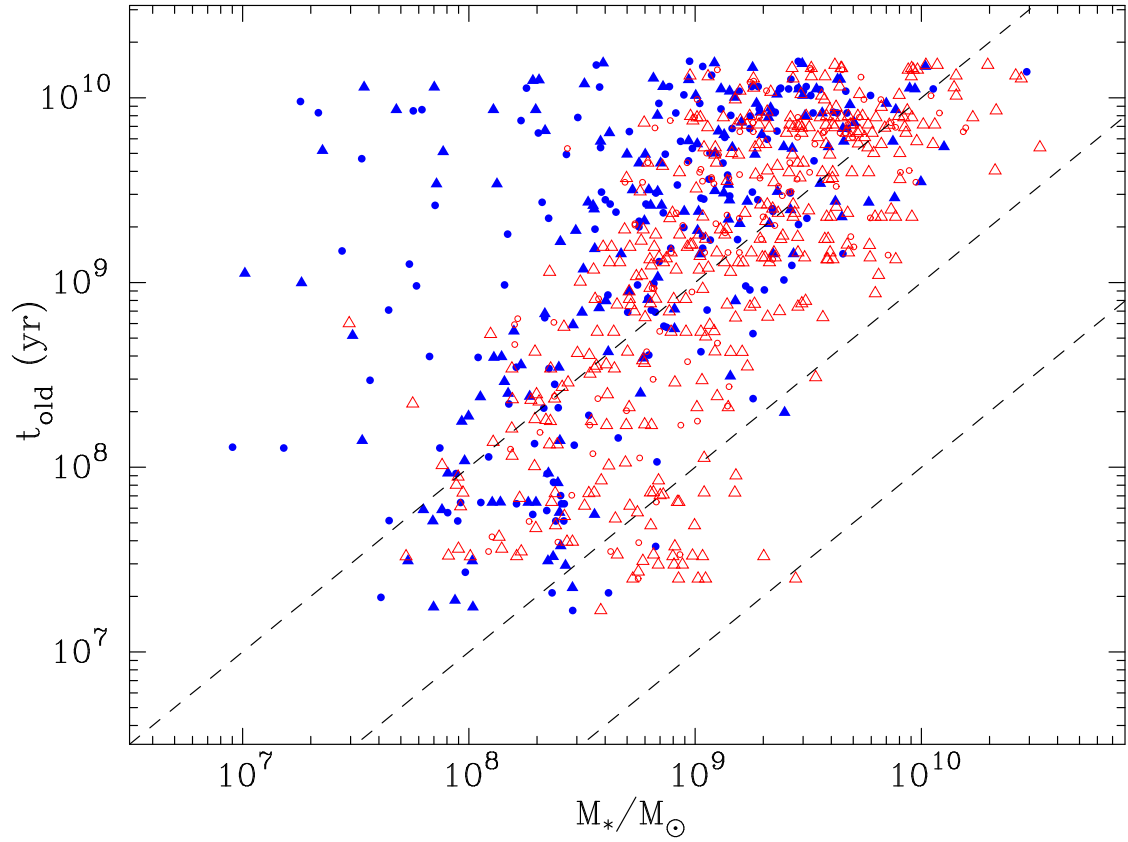}} 
\figcaption{Age $t$(old) of the oldest stellar population
in the galaxy vs. its total stellar mass $M_*$. 
The dashed lines from left to right correspond to loci with 
the SFR = 1, 10 and 100 $M_\odot$ yr$^{-1}$, respectively. \label{fig16}}
\end{figure*}

Fig. \ref{fig16} shows how the age of the oldest stellar
population $t$(old) depends on 
the total stellar mass $M_*$. The dashed lines 
are the loci of constant continuous star formation rate 
corresponding to, from left to right,  
SFR = 1, 10 and 100 $M_\odot$ yr$^{-1}$. 

It can be seen that galaxies with high EW(H$\beta$) ($\ge$ 100\AA) 
denoted by filled symbols do not show any evident trend. 
On the other hand, there is an increase of 
$t$(old) with increasing $M_*$ for galaxies with 
EW(H$\beta$) $<$ 100\AA. This implies that, among the galaxies with low 
EW(H$\beta$), massive galaxies are in general older than lower mass 
galaxies. 
The absence of such a dependence for galaxies with high EW(H$\beta$)
may in part be explained by the uncertainties in the SED modeling. Indeed,
the fraction of the light produced by the old stellar population in spectra
of galaxies with a high EW(H$\beta$) is much lower than that in galaxies
with a low EW(H$\beta$), both because of the brighter young stellar population
and of the higher contribution of gaseous emission. Furthermore, galaxies
with high EW(H$\beta$) have systematically higher redshifts compared
to galaxies with low EW(H$\beta$). Their 
larger distances make their continua weaker, which renders
 their SED fitting more uncertain.
This is because of the 
following selection effect.
We select only galaxies with [O {\sc iii}] $\lambda$4363 that can be measured
with good accuracy for abundance determination, and this line is stronger
in galaxies with high EW(H$\beta$). 

The location of the galaxies in Fig. \ref{fig16} relative to the dashed lines indicates continuous SFRs that are somewhat lower 
than the ongoing SFRs derived by equations \ref{eq3} and \ref{eq4}, and shown
 in Fig. \ref{fig15}a. The largest offset is
for low-mass galaxies with high EW(H$\beta$) (filled symbols). 
The vast majority of these galaxies are located in the
region corresponding to continuous SFRs $<$ 1 $M_\odot$ yr$^{-1}$, i.e. 
$\ga$ 10 times lower
than those derived from the ongoing star formation episode. 
This discrepancy is again 
an indication that star formation in the LCGs  
occurs in rare bursts of short duration, so that the SFR averaged over 
the lifetime of the galaxy is more than one order of magnitude lower
than the one derived from the H$\alpha$ luminosity. The
bursting nature of star formation is less pronounced for the 
higher-mass galaxies with low
EW(H$\beta$) (open symbols): their continuous SFRs averaged over the galaxy's
lifetime are only 
$\sim$ 2 - 3 times lower than those derived from the ongoing 
episode of star formation. 

\section{Conclusions}

We discuss here the properties of a sample of 
luminous compact star-forming galaxies (LCG) 
selected 
from the Data Release 7 (DR7) of the Sloan Digital Sky Survey (SDSS).
We show that the so-called ``green pea'' galaxies discussed 
by  \citet{C09} are a subset of these LCGs. However, in contrast to 
\citet{C09} who selected their green pea galaxies solely 
on the basis of broad-band
SDSS colors and compact structure, our LCG galaxies were chosen also 
on the basis of their spectroscopic properties. 
Our main selection criteria are as 
follows: a) a high extinction-corrected luminosity 
$L$(H$\beta$) $>$ 3$\times$10$^{40}$ erg s$^{-1}$ of the H$\beta$ emission
line; b) a high equivalent width EW(H$\beta$) $\ge$ 50\AA\ of the H$\beta$
emission line; c) a strong [O {\sc iii}] $\lambda$4363 emission line allowing
accurate abundance determination; d) a compact structure on SDSS images;
and e) an absence of obvious AGN spectroscopic features. 
Using these criteria, we 
selected in total 803 galaxies, split into four subsamples based on the
value of EW(H$\beta$) (greater or smaller than 100\AA) and morphological
appearance (round shape or some evidence of disturbed morphology). 

Analysis of the LCG sample has led to the following findings:

1. The selected galaxies have redshifts between 0.02  and 0.63, a redshift 
range that is 
a factor $\ga$ 2 higher than the redshift range $z$ = 0.112 -- 0.360 
of the green pea sample of \citet{C09}. 
We find that the properties of LCGs and green pea galaxies selected
from the SDSS are in many respects similar to those of well-studied nearby 
luminous blue compact dwarf (BCD) galaxies, such as SBS 0335--052E and
HS 0837+4717.
Since the color of an emission-line 
galaxy on the $gri$ SDSS composite 
image changes with its redshift, depending on which strong 
emission lines fall within a particular filter band,  
 the LCGs display  
a variety of colors from intense blue for nearby galaxies with $z$ $<$ 0.07,
through pink and green at redshifts $z$ $\sim$ 0.09 and $\sim$ 0.15 -- 0.30,
respectively, to brown at redshifts $z$ $\sim$ 0.6. Therefore, the green pea 
galaxy sample is just a subsample with a narrow redshift range 
of a larger LCG sample.
A significant fraction of our LCGs have low redshifts
($z$ $<$ 0.1), thus allowing the study of LCG properties in more 
detail than in the previous investigations of \citet{C09} and \citet{A10}.

2. Although our selection criteria 
do not set an upper limit on the luminosity $L$(H$\beta$) of
the H$\beta$ emission line, we find that there is
an upper bound to $L$(H$\beta$) at $\sim$ 3$\times$10$^{42}$ erg s$^{-1}$.  
This implies a self-regulating mechanism of star formation that does not allow
the formation of star-forming regions brighter than that limit in a galaxy.

3. In the [O {\sc iii}] $\lambda$5007/H$\beta$ vs
[N {\sc ii}] $\lambda$6583/H$\alpha$ diagnostic diagram, 
LCGs occupy the region of star-forming galaxies with the highest 
excitation. On the other hand, the five 
low-metallicity AGNs discovered by \citet{I07,I10} and \citet{IT08}
also lie in the same region. This implies that 
further detailed studies of LCGs may result in the discovery of 
a larger number of 
low-metallicity AGNs showing some Seyfert 1 features such as broad hydrogen 
emission lines.

4. We find that the oxygen abundances 12 + log O/H in LCGs are in the
range 7.6 -- 8.4 with a median value of $\sim$ 8.11, confirming the results
of \citet{A10} concerning a subset of the green pea sample of \citet{C09}.
This range of oxygen abundances is typical of nearby 
lower-luminosity BCDs. We note, however, that
the median oxygen abundance of $\sim$ 8.7, derived by \citet{C09} for 
their green pea sample is overestimated, 
because those authors used empirical relations for abundance 
determinations, instead of the direct method
used in this paper and in \citet{A10}. There is no dependence of
12 + logO/H on redshift. On the other hand, we find that the oxygen 
abundances in galaxies with higher equivalent width EW(H$\beta$) are 
generally lower. As for the 
N/O, Ne/O, S/O, Cl/O, Ar/O and Fe/O abundance ratios, they are similar to
those found for lower-luminosity BCDs. Overall, we find 
no appreciable difference in element abundances between LCGs and BCDs, implying
a similar chemical enrichment history.

5. In the luminosity-metallicity diagram, LCGs are shifted by $\sim$ 2 mag
to brighter magnitudes as compared to the bulk of nearby emission-line 
galaxies. However, together with the most metal-deficient nearby BCDs I Zw 18,
SBS 0335--052W, SBS 0335--052E, extremely metal-poor intermediate-redshift
galaxies \citep[$z$ $<$ 1, ][]{K07}, luminous compact metal-poor galaxies
\citep[$z$ $\sim$ 0.7, ][]{H05} and Lyman-break galaxies (LBG)
\citep[$z$ $\sim$ 3, ][]{P01}, LCGs form a common luminosity-metallicity 
relation, that for the most actively star-forming galaxies. 
Some LCGs have oxygen abundances
and luminosities similar to LBGs, despite much lower redshifts. This opens 
the possibility of studying LBGs through brighter and hence easier to 
observe LCGs.

6. We develop a technique for determing galaxy masses by 
fitting the observed spectral energy distribution (SED) by a model SED 
that includes both emission from 
stellar populations of different ages and the continuum emission 
from the ionized gas, the latter having not been considered by 
previous investigators. We show that such an omission results in an 
overestimate of the stellar mass. This effect is especially 
strong for galaxies with high EW(H$\beta$) $>$ 100\AA. Thus, the median
mass of the galaxies in our sample is a factor of $\sim$ 3 times lower
than that derived by \citet{C09} and used by \citet{A10}. 

7. The total stellar mass of LCGs is in the range $\sim$ 10$^7$ -- $\sim$
10$^{11}$ $M_\odot$, with a median value of $\sim$ 10$^9$ $M_\odot$.
The median mass of $\sim$ 3$\times$10$^7$ $M_\odot$ of the young stellar 
population is a factor of $\sim$ 30 lower. We find that the mass
of the young stellar population increases with increasing total stellar
mass. However, the fraction  by mass of the young stellar population is
higher in galaxies with a low total stellar mass. 
Furthermore, H {\sc ii} regions
in low-mass LCGs are systematically hotter and denser, 
than those in high-mass galaxies.

8. We find a clear increase of the oxygen abundance with increasing total 
stellar mass of the galaxy in the mass-metallicity diagram. 
However, there is
no correlation between the N/O abundance ratio and the 
total stellar mass for the lowest-mass galaxies, those with  
EW(H$\beta$) $\ge$ 100\AA, implying primary N production in those objects.
There is however a total mass dependence of N/O for the highest-mass 
galaxies, those with  EW(H$\beta$) $<$ 100\AA, implying some 
secondary of N in those galaxies. 

9. The star formation rate in LCGs varies in a large range $\sim$ 0.7 --
$\sim$ 60 $M_\odot$ yr$^{-1}$ with a median value of 
$\sim$ 4 $M_\odot$ yr$^{-1}$. The range and median value of the SFRs in LCGs 
is lower by a factor of several, as compared to  
Lyman-break galaxies at redshift $\sim$ 3 \citep{P01} and 
higher-redshift galaxies \citep{S09,SB10}. However, the highest SFRs found in 
LCGs compare well with those in high-redshift galaxies. LCGs have 
specific star formation rates that are among the highest
 known, in the range
$\sim$ 10$^{-7}$ -- 10$^{-9}$ yr$^{-1}$, and comparable 
to those of high-redshift galaxies
in the $z$ $\sim$ 3 -- 8 redshift range. 
This similarity offers the possibility of studying 
extreme star formation in high-redshift galaxies by
using LCGs as relatively nearby proxies.

10. We find that the age of the oldest stellar
population increases with increasing stellar mass in the LCGs with low 
EW(H$\beta$) ($<$ 100\AA). However, no such dependence is found
for galaxies with high EW(H$\beta$), an absence which can 
be explained by larger
uncertainties in fitting the spectra of these galaxies. 
Star formation in LCGs at the rate derived from H$\alpha$ emission 
cannot be continuous, but rather must occur in strong short bursts 
separated by long quiescent phases, just as in BCDs. 
 The bursting nature of star formation is more
pronounced in LCGs with higher EW(H$\beta$) and lower masses.

\acknowledgements
Y.I.I. is grateful to the staff of the Astronomy Department at the 
University of Virginia for their warm hospitality. 
Support for this work is provided by  NASA. 
Y.I.I. and N.G.G. acknowledge the support of the Cosmomicrophysics project of 
the National Academy of Sciences of Ukraine.
Funding for the Sloan Digital Sky Survey (SDSS) and SDSS-II has been 
provided by the Alfred P. Sloan Foundation, the Participating Institutions, 
the National Science Foundation, the U.S. Department of Energy, the National 
Aeronautics and Space Administration, the Japanese Monbukagakusho, and the 
Max Planck Society, and the Higher Education Funding Council for England.





  \begin{deluxetable}{ccccccc}
  \tablecolumns{7}
  \tablewidth{0pc}
  \tablecaption{Average characteristics of LCGs \label{tab1}}
  \tablehead{
  \colhead{Sample}&\colhead{$z$}&\colhead{$g$}&\colhead{$E$($B-V$)}&\colhead{$M$$_{\rm total}$}&\colhead{$M$$_{\rm young}$}
&\colhead{$t$$_{\rm young}$}\\
  \colhead{}&\colhead{}&\colhead{}&\colhead{}&\colhead{10$^9$$M_\odot$}&\colhead{10$^7$$M_\odot$}
&\colhead{Myr}
}
  \startdata
1    &0.201&20.3&0.10& 0.68 & 2.87 &3.0 \\
2    &0.129&19.3&0.12& 0.64 & 2.37 &3.3 \\
3    &0.152&19.2&0.12& 1.31 & 3.86 &4.4 \\
4    &0.094&18.4&0.15& 1.39 & 3.62 &4.0 \\ \tableline
Total&0.116&18.9&0.13& 1.07 & 3.12 &3.6 \\ \tableline \tableline
{Sample}&{$T_e$(O {\sc iii})}&{$N_e$(S {\sc ii})} &{12+logO/H}&{SFR(H$\alpha$)}&{SFR(FUV)}\\
        &      K             &     cm$^{-3}$      &           &{$M_\odot$ yr$^{-1}$}&{$M_\odot$ yr$^{-1}$} \\ \tableline
1    & 13100 & 180&8.04&5.0&3.4 \\
2    & 12400 & 130&8.09&4.1&3.3 \\
3    & 12200 & 120&8.09&4.3&4.0 \\
4    & 11700 &  90&8.16&3.4&4.5 \\ \tableline
Total& 12200 & 110&8.11&4.1&3.9 \\
  \enddata
  \end{deluxetable}

\clearpage

  \begin{deluxetable}{crrcrccr}
  \tabletypesize{\footnotesize}
  \tablecolumns{7}
  \tablewidth{0pc}
  \tablecaption{Characteristics of \citet{C09} Green Pea galaxies
included in the LCG sample \label{tab2}}
  \tablehead{
  \colhead{SDSS ID}&\colhead{RA}&\colhead{Dec}&\multicolumn{2}{c}{\citet{C09}}
&&\multicolumn{2}{c}{this paper}\\ \cline{4-5} \cline{7-8}
  \colhead{}&\colhead{J2000}&\colhead{J2000}&\colhead{12+logO/H}
&\colhead{log($M_*$/$M_\odot$)}&\colhead{}&\colhead{12+logO/H}&\colhead{log($M_*$/$M_\odot$)}
}
  \startdata
588848899919446344& 195.546460& $-$0.087897&8.75&  9.08&& 8.12&  9.02 \\
587725576962244831& 261.776373&   59.817273&8.71&  9.81&& 8.24&  9.26 \\
587731187273892048& 351.413453&    0.752012&8.70&  9.38&& 8.29&  9.41 \\
587727179006148758&  45.839226& $-$7.989791&8.71&  8.75&& 7.86&  9.15 \\
587724240158589061&  54.949128& $-$7.428132&8.78&  9.75&& 8.38&  9.70 \\
587726032253419628& 191.097382&    2.261231&8.70&  9.45&& 8.17&  9.65 \\
588010360138367359& 130.570630&    3.635203&8.55&  9.71&& 8.19&  9.38 \\
587726102030451047& 236.787938&    3.603914&8.74&  8.96&& 8.16&  8.79 \\
587729155743875234& 173.265848&   65.228162&8.66& 10.05&& 7.97&  9.30 \\
587728919520608387& 212.938906&   62.653138&8.67& 10.18&& 8.20&  9.09 \\
587729229297090692& 234.405309&   58.794575&8.58&  9.27&& 8.04&  9.16 \\
587730774416883967& 339.396081&   13.613062&8.55& 10.10&& 8.16&  9.45 \\
587730774965354630&   6.716985&   15.460460&8.81&  9.58&& 8.07&  9.02 \\
587728906099687546& 117.403215&   33.621219&8.79&  9.85&& 8.29&  9.49 \\
587725550133444775& 156.563375&   63.552363&\nodata&  9.05&& 8.09&  8.23 \\ 
588009371762098262& 170.582224&   61.912629&8.70&  8.71&& 8.29&  7.85 \\
588011122502336742& 181.772142&   61.586621&\nodata&  9.85&& 8.09&  8.38 \\
588013384341913605& 141.501678&   44.460044&8.62&  9.08&& 8.01&  8.78 \\
587732134315425958& 195.368010&   51.080893&8.63&  9.53&& 8.00&  9.98 \\
587729777439801619& 204.299529& $-$2.434842&8.76&  9.94&& 8.26&  9.19 \\
587729777446945029& 220.630713& $-$2.164466&8.57&  8.80&& 7.94&  8.65 \\
587732152555864324& 116.991682&   23.609113&\nodata&  9.40&& 7.98&  9.06 \\
587732578845786234& 157.912214&    7.265701&8.52&  9.02&& 8.33&  8.75 \\ 
587733080270569500& 163.378431&   52.631353&8.78&  9.75&& 8.10&  9.77 \\
588297864714387604& 131.975356&   33.615227&8.81&  9.50&& 8.02&  9.37 \\
587735696987717870& 213.630037&   54.515587&8.58&  8.81&& 7.94&  8.02 \\
587733441055359356& 251.527242&   31.514859&\nodata&  8.76&& 8.03&  9.07 \\
588017605211390138& 154.513517&   41.105860&8.68&  9.82&& 7.82&  9.32 \\
588017114517536797& 216.023868&   42.279524&8.78&  9.00&& 8.04&  8.34 \\
588017116132540589& 228.535985&   38.868716&\nodata&  8.90&& 8.40&  9.32 \\
588018090541842668& 235.755108&   34.767079&\nodata&  8.76&& 8.00&  8.05 \\
588018090013098618& 251.898063&   22.783002&8.61&  9.27&& 8.06&  8.76 \\
588016878295515268& 137.879799&   31.457439&\nodata&  9.27&& 7.94&  8.37 \\
587735661007863875& 139.260529&   31.872384&8.95&  9.44&& 8.34&  9.37 \\
588018055114784812& 220.041419&   46.326930&8.72&  9.83&& 8.09&  9.62 \\
588018055652769997& 223.648271&   45.482288&8.84& 10.22&& 8.11&  9.52 \\
588017570848768137& 192.144310&   12.567480&8.53&  9.06&& 8.11&  9.05 \\
587736915687964980& 241.152768&    8.333082&8.75&  9.90&& 8.02&  8.48 \\
587736915687375248& 239.858241&    8.688655&8.64&  8.74&& 8.07&  8.97 \\
587738410863493299& 152.987850&   13.139471&8.64&  8.66&& 8.01&  8.31 \\
587735349111947338& 184.766599&   15.435698&8.82&  8.66&& 7.89&  8.35 \\
587738570859413642& 204.867933&   15.278369&8.83&  9.96&& 8.09&  9.43 \\
587736940372361382& 217.614622&   34.154720&\nodata&  9.31&& 7.99&  8.75 \\
587739153352229578& 117.990764&   16.637010&8.65&  9.96&& 7.87&  8.35 \\
587738947196944678& 123.966679&   21.939902&8.55&  8.72&& 8.00&  8.71 \\
587738371672178952& 125.698590&   22.695578&8.81&  9.33&& 8.08&  8.43 \\
588017978880950451& 150.556494&   34.704908&8.64&  9.58&& 8.01&  8.52 \\
587739408388980778& 174.342249&   35.407413&8.76&  9.32&& 8.16&  9.56 \\
588017977277874181& 171.657352&   38.050810&8.56&  9.79&& 8.04&  8.83 \\
587739406242742472& 178.020352&   34.013853&\nodata&  8.86&& 7.96&  8.35 \\
587739828742389914& 224.396405&   22.533833&8.55&  9.30&& 8.04&  9.13 \\
587739652107600089& 238.041673&   21.053410&8.68&  9.90&& 7.91&  9.46 \\
587739721387409964& 249.330431&   14.651378&8.62&  9.64&& 8.30&  9.41 \\
587741600420003946& 181.252807&   26.346595&8.81&  9.66&& 8.23&  9.84 \\
587741421099286852& 126.715863&   18.347732&8.75&  8.67&& 8.23&  8.81 \\
587741532770074773& 133.350354&   19.506280&8.68&  9.37&& 8.13&  9.32 \\
587741817851084830& 137.805603&   18.518936&8.89&  9.93&& 8.00&  9.75 \\
587741392649781464& 152.329151&   29.272638&8.69&  8.55&& 7.91&  7.87 \\
587741490367889543& 158.112322&   27.298680&8.69& 10.02&& 8.22&  9.65 \\
587742014876745993& 141.869487&   17.671838&8.51&  9.38&& 8.01&  9.26 \\
588023240745943289& 140.705287&   19.227629&8.59& 10.48&& 8.30&  9.56 \\
587745243087372534& 141.384863&   14.053623&8.73&  9.08&& 7.94&  8.46 \\
587744874785145599& 121.325174&    9.425978&\nodata& 10.24&& 8.28&  9.36 \\
587742013825941802& 197.653081&   21.804731&8.83&  9.27&& 8.54&  9.46 \\
587742062151467120& 196.734804&   22.694003&\nodata& 10.10&& 8.02&  9.82 \\
587741727655919734& 193.761316&   25.935911&\nodata& 10.02&& 8.10&  9.45 \\ 
  \enddata
  \end{deluxetable}

\clearpage

\begin{thebibliography}{}

\bibitem[Abazajian et al. (2009)]{A09} Abazajian K. N., et al., 
2009, \apjs, 182, 543



\bibitem[Aller (1984)]{A84} Aller, L. H. 1984, Physics of Thermal Gaseous 
Nebulae (Dordrecht: Reidel) 

\bibitem[Amor\'in et al. (2010)]{A10} Amor\'in, R. O., P\'erez-Montero, E., \& 
V\'ilchez, J. M. 2010, \apj, 715, L128

\bibitem[Asplund et al. (2009)]{As09} Asplund, M., Grevesse, N., Sauval, A.J., 
\& Scott, P. 2009, ARA\&A, 47, 481

\bibitem[Baldwin et al. (1981)]{B81} Baldwin, J. A., Phillips, M. M., 
\& Terlevich, R. 1981, \pasp, 93, 5

\bibitem[Cardamone et al. (2009)]{C09} Cardamone, C., et al. 2009, MNRAS, 
399, 1199

\bibitem[Cardelli et al. (1989)]{C89} Cardelli, J. A., Clayton, G. C., \&
Mathis, J. S. 1989, \apj, 345, 245

\bibitem[Fioc \& Rocca-Volmerange (1997)]{FR97} Fioc, M., \& 
Rocca-Volmerange, B. 1997, \aap, 326, 950

\bibitem[Girardi et al. (2000)]{Gi00} Girardi, L., Bressan, A., Bertelli, G., 
\& Chiosi, C. 2000, \aaps, 141, 371

\bibitem[Gonz\'alez Delgado et al. (2005)]{G05} Gonz\'alez Delgado, R. M., 
Cervi\~no, M., Martins, L. P., Leitherer, C., \& Hauschildt, P. H. 
2005, \mnras, 357, 945


\bibitem[Guseva et al. (2006)]{G06} Guseva, N. G., Izotov, Y. I., 
\& Thuan, T. X. 2006, \apj, 644, 890

\bibitem[Guseva et al. (2007)]{G07} Guseva, N. G., Izotov, Y. I., 
Papaderos, P., \& Fricke, K. J. 2007, \aap, 464, 885

\bibitem[Guseva et al. (2009)]{G09} Guseva, N. G., Papaderos, P., Meyer, H. T.,
Izotov, Y. I., \& Fricke, K. J. 2009, \aap, 505, 63

\bibitem[Hoyos et al. (2005)]{H05} Hoyos, C., Koo, D. C., Phillips, A. C., 
Willmer, C. N. A., \& Guhathakurta, P. 2005, \apj, 635, L21

\bibitem[Izotov \& Thuan (2008)]{IT08} Izotov, Y. I., \& Thuan, T. X. 
2008, \apj, 687, 133

\bibitem[Izotov \& Thuan (2004)]{IT04} Izotov, Y. I., \& Thuan, T. X. 
2004, \apj, 602, 200

\bibitem[Izotov et al. (1990)]{I90} Izotov, Y. I., Guseva, N. G., 
Lipovetsky, V. A., Kniazev, A. Y.., \& Stepanian, J. A. 1990 \nat, 343, 238

\bibitem[Izotov et al. (1994)]{I94} Izotov, Y. I., Thuan, T. X., 
\& Lipovetsky, V. A. 1994, \apj, 435, 647



\bibitem[Izotov et al. (1997)]{I97} Izotov, Y. I., Lipovetsky, V. A., 
Chaffee, F. H., Foltz, C. B., Guseva, N. G., \& Kniazev, A. Y. 1997, 
\apj, 476, 698

\bibitem[Izotov et al. (2006)]{I06} Izotov, Y. I., Stasi\'nska, G., 
Meynet, G., Guseva, N. G., \& Thuan T. X. 2006, \aap, 448, 955

\bibitem[Izotov et al. (2007)]{I07} Izotov, Y. I., Thuan, T. X., \&
Guseva, N. G. 2007, \apj, 671, 1297

\bibitem[Izotov et al. (2010)]{I10} Izotov, Y. I., Guseva, N. G., 
Fricke, K. J., Stasi\'nska, G., Henkel, C., \& Papaderos, P. 2010, \aap,
in press; preprint arXiv:1005.1844

\bibitem[Kakazu et al. (2007)]{K07} Kakazu, Y., Cowie, L. L., \& Hu, E. M. 
2007, ApJ, 668, 853

\bibitem[Kauffmann et al. (2003)]{K03} Kauffmann, G., Heckman, T. M., 
Tremonti, C., et al. 2003, \mnras, 346, 1055

\bibitem[Kennicutt (1998)]{K98} Kennicutt, R. C., Jr. 1998 \araa, 36, 189



\bibitem[Lintott et al. (2008)]{L08} Lintott, C. J., Schawinski, K., 
Slosar, A., et al. 2008, \mnras, 389, 1179

\bibitem[Lintott et al. (2011)]{L11} Lintott, C. J., Schawinski, K., 
Bamford, S., et al. 2011, \mnras, 410, 166

\bibitem[Pettini et al. (2001)]{P01} Pettini, M., Shapley, A. E., 
Steidel, C. C., et al. 2001, \apj, 554, 981

\bibitem[Pustilnik et al. (2004)]{P04} Pustilnik, S., Kniazev, A., 
Pramskij, A., Izotov, Y., Foltz, C., Brosch, N., Martin, J.-M., \& 
Ugryumov, A. 2004, \aap, 419, 469


\bibitem[Schaerer \& de Barros (2010)]{SB10} Schaerer, D., \& de Barros, S. 
2010, \aap, 515, 73

\bibitem[Schneider et al. (2010)]{S10} Schneider, D. P., et al. 2010, \aj,
139, 2360

\bibitem[Stark et al. (2009)]{S09} Stark, D. P., Ellis, R. S., Bunker, A., 
Bundy, K., Targett, T., Benson, A., \& Lacy, M. 2009, \apj, 697, 1493

\bibitem[Stasi\'nska et al. (2006)]{S06} Stasi\'nska G., Cid Fernandes R., 
Mateus A., Sodr/'e L., \& Asari N. V., 2006, mnras, 371, 972

\bibitem[Thuan (2008)]{T08} Thuan, T. X. 2008, in 
Low-metallicity star formation: from the first stars to dwarf galaxies,
ed. L. Hunt, S.C. Madden, \& R. Schneider (Cambridge: Cambridge Univ. 
Press), 348

\bibitem[Thuan \& Izotov (2005)]{TI05} Thuan, T. X., \& Izotov, Y. I. 
2005, \apjs, 161, 240

\bibitem[Thuan et al. (1997)]{T97} Thuan, T. X., Izotov, Y. I., \& 
Lipovetsky, V. A. 1997, \apj, 477, 661

\bibitem[Tremonti et al. (2004)]{T04} Tremonti, C. A., Heckman, T. M., 
Kauffmann, G., et al. 2004, \apj, 613, 898

\bibitem[Whitford (1958)]{W58} Whitford, A. E. 1958, \aj, 63, 201

\end{thebibliography}
\end{document}